\documentclass[a4paper, astronm]{article} %

\RequirePackage{graphicx}
\RequirePackage{pifont,latexsym,ifthen,theorem,rotating,calc,textcase,booktabs,color}
\RequirePackage{amsfonts,amssymb,amsbsy,amsmath}
\RequirePackage[errorshow]{tracefnt}
 
\usepackage{footnote}
\usepackage{epstopdf}
\usepackage{astronm}
 
\usepackage{float}
\usepackage[caption = false]{subfig}

\newcommand{\ud}{\mathrm{d}}

\title{ Dynamic Modelling of Health and its application to the large scale analysis of Body Mass Index, using data from  consecutive set of surveys.
}
 
\author{Vladislav Moltchanov  \\
\\
National Institute for Health and Welfare (THL)\\
Department of Health \\
P.O. Box 30
FI-00271 Helsinki
Finland \\
\emph{email}: vladislav.moltchanov@thl.fi }
 
\begin{document}
\maketitle
 
\begin{abstract} 
The methods used so far for the analysis of time changes in population  health
suffer from the lack of causality in their design.   This results in problems   with their implementation  and interpretation.  Here the method is presented with causality directly implemented in the design.  This is done by,  first,  building up a dynamic model  of population,  postulating  existence  of   Driving Force acting at subjects, while they move along their cohort lines,  causing the  changes  of their substantial  health  indicators , State Variables, at rate proportional  to this Force.  The correspondent   rates ,  named  Cohort Trends,  or C-trends,  describe health history  in each birth cohort.      Having initial value   and  C-trends  ,  the model  allows to calculate health  level  (the means of  State Variables) in each birth cohort, and  thus,  in the whole population.
The task for statistical method  is to  identify  the dynamic model  (evaluate C-trends and Initial  values)  using data from a set of consecutive  independent cross-sectional surveys.  This is done by an iterative  algorithm, running  multiple regression procedure at each step, until the specified smoothing conditions are fulfilled.   The algorithm can operate with   surveys  having  different age ranges.
The illustrative example  shows  the results of analysis of  Body  Mass Index  for men , using 7 surveys  in  period  1972-2002  with age ranges  25-64 and  25-74.
Since  C-Trend   is  proxy  for  Driving Force,  the  year - age pattern of C-Trends  provides unbiased  information  for health authorities   on  efficiency of  health promotion  actions or negative effects  of uncontrolled  harmful   factors.
\end{abstract}

 

\section{Introduction}
\paragraph{}
This paper deals with the principles and methods of the analysis of  time changes in population health, based on  dynamic  modeling of population.
 
The first step of analysis is  building up a dynamic model  of population,  postulating  existence  of   Driving Force acting at subjects, while they move along their cohort lines,  causing changes  of their substantial  health  indicators, State Variables, at rate proportional  to this Force.  The correspondent   rates,  named  Cohort Trends,  or C-trends,  describe health history  in each birth cohort.   Having initial values   and  C-trends  for each birth cohort,  the model  allows to calculate health  level  (means of  State Variables) in each birth cohort, and  thus,  in the whole population.
 
In dynamic view, all traditional health indicators fall into two main categories: Driving Force (this includes smoking, physical activities, diets habits)   and  State Variables (among those BMI, Blood pressure, Cholesterol levels). The third category includes all class indicators, like gender, social group, place of residence.
 
Driving Force and State Variable indicators play different role in dynamic model and should be analyzed using different schemes.
 
In this paper we focus on analysis of State variable BMI. The data for this variable are taken from a set
of  independent consecutive  health surveys,  with randomly selected samples  of  population.  Age ranges of samples and  time periods between surveys  could be different.
Total calendar period of analysis could be quite large, in the example  presented  here,  it is 30 years, covering  7  health surveys.
 
The methods used so far, addressing  time changes in population health, did not differentiate between Driving Force and State Variable indicators.
The  problem has been approached by assessing  trends over
time for means  and other statistics (e.g. percentiles) of the
age-categories-specific distributions of parameters of interest,
such as traditional risk factor indicators (e.g. systolic and
diastolic blood pressure, cholesterol, body mass index), their
categorical derivatives  (e.g. prevalence of high blood pressure,
prevalence of high cholesterol, prevalence of obesity) and event and
mortality rates, see, for example,
\cite{MoniMono,Dobson1988a,Dobson1988b,Kuulasmaa2000,Edward2005,Kautiainen2002,Chen2003,DiLiberti2001}.
 
Various terms are used in  literature for such trends: "trends"
\cite{MoniMono,Dobson1988a,Dobson1988b,Kuulasmaa2000}, "secular
trends" \cite{Edward2005,Kautiainen2002}, "time trends"
\cite{Holford1991}. Here we will call  them  "secular trends".
There were several modifications in methods used to assess secular
trends. One of these,  "trends by linear regression",  was the key
element of  the analysis in the WHO MONICA Project
\cite{MoniMono,Dobson1988a,Kuulasmaa2000}. \cite{MoniMono} contains
extended list of publications, the survey-specific tabulations could
be found in \cite{surveydb}. First, the age-group specific trends
were assessed, using linear regression; then they were aggregated
using direct age standardization with fixed weights. The aggregated
trends were subject to correlation analysis in order to test MONICA
hypotheses.
 
A different modification uses the multiple logistic regression
procedure  applied to the whole set of data \cite{Edward2005}. As
a result, the marginal characteristics are obtained directly from
the procedure. This is equivalent to direct standardization, with
weights corresponding  to the analyzed population.
 
In previous two examples, the method was applied to the data of wide
age-range (40 years and more), while spans  between consecutive
surveys were 3-10  years. Some studies of adolescents deal with
samples of age range  5-8 years, being  sampled every year or every
other year. In that case the trends are first  examined visually by
age \cite {Kautiainen2002,Chen2003}, since, as a rule, they exhibit
a wide  diversity of age-specific  patterns.
 
To cope with such a diversity, several approaches have been commonly used.
The first one is the APC ( Age, Period, Cohort) adjustment
\cite{Selvin1996,Holford1991}, which accounts for high variability of
secular trends across age, time period and cohorts
\cite{Chen2003}
The second one divides the overall time period into several segments
\cite {DiLiberti2001}, for which the corresponding plots suggest
linear trends. The third one first aggregates the age-specific
values, using  direct age standardization, then estimates trends by
applying linear regression to the aggregated values \cite
{Kautiainen2002}.
 
The above examples bring to the fore the following fact: in many
cases the plots of the analyzed parameters versus time display wide
diversity across age-groups, and do not suggest linear trends within
age groups. Even if the analysis is applied to the data comprising
only two surveys \cite{Dobson1988a,Dobson1988b}, the unmeasured
data between these two surveys may exhibit high nonlinearity.
 
Various types of summary rates,  have been criticized, for example,
by Holford \cite{Holford1991}: "Although this approach has the
advantage of simplicity, it suffers from significant limitations;
important details in the trends are lost in the averaging   process
involved in generating a summary rate. In many instances, these
details have contributed significantly to the understanding of time
trends for disease". However, the APC adjustment suggested by him
\cite{Holford1991} and Selvin  \cite{Selvin1996}, as an alternative
to summary rates, is also not free of problems. The main one stems
from the fact that parameters used for adjustment "are hopelessly
entangled" \cite{Holford1991}.
 
A variety of methods have been suggested recently to solve the problem of identifiability in the APC analysis.
One of them, Partial Least Square  (PLS), is illustrated in \cite{Jiang} in analysis of BMI.
\cite{Havulinna}  claims that model identifiability is becoming less of a problem with Bayesian APC models, and suggests two extension. one involving interactions
between the age, period and cohort effects. Second extension uses autoregressive integrated (ARI)
models.
 
At the same time, in review of the latest methods  of Age-Period-Cohort Analysis, it was acknowledged  that  "Although a variety of approaches has been proposed to solve the APC conundrum,  each has limitations.
Yet another challenge is a criticism often lodged against general -purpose methods of APC analysis, namely, they provide no venue for testing specific, substantive, and mechanism-based hypotheses and thus are mere accounting devices of algebraic convenience that may be misleading."  \cite[p3]{Yang}.
 
\begin{figure}
\begin{center}
\centerline{\includegraphics[angle=90,width=20.5cm]{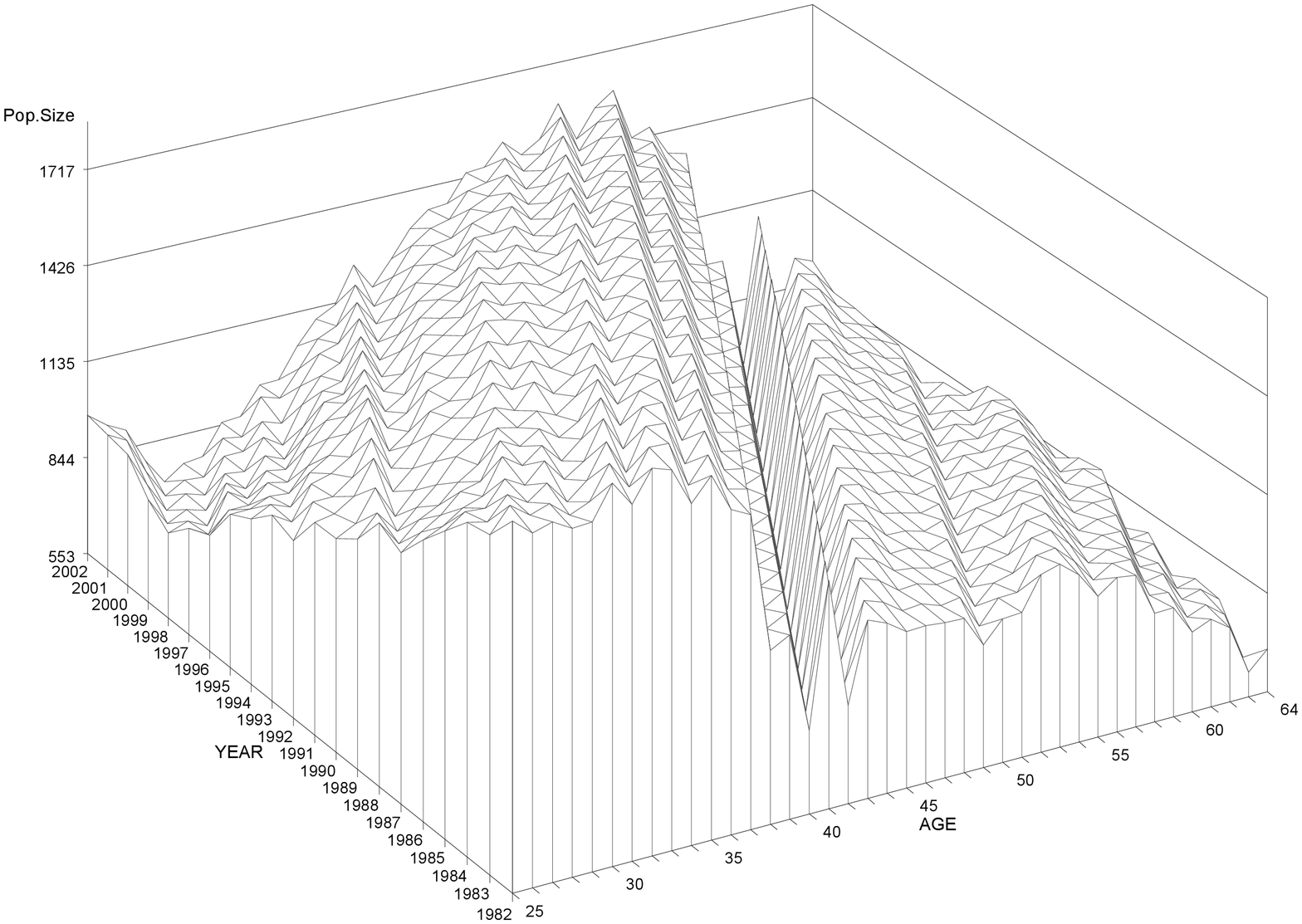}}
\end{center}
\caption{Example of Population size, Men. Population register counts by year and age.
\label{popsize}}
\end{figure}

According to \cite{Holford1991}, secular trends "are significant
because they can be highly suggestive as to what might be expected
in the future and they are an effective approach to understanding
disease etiology". We believe, that  in many cases, the real data reject the idea
of such a trends. For example, the real data on population size shown in
Figure ~\ref{popsize} hardly suggest age-specific linear trends  over time.
Rather they show fairly smooth changing along cohort lines.
 
To develop dynamic approach, a certain  knowledge and skills are required not only in
statistics, but also in physics, theory of control, theory
of optimization, differential equations and others \cite{Luenberger}. Perhaps, this is the reason why  the
dynamic systems approach so far has not been widely used in
population health research, though some  examples of such
applications can be found in the literature (see, for example,
\cite{Harg}).
 
General purpose of the dynamic system applications  is formulated in \cite{BrownUni}: "These three categories correspond roughly to the need to predict,
explain, and understand physical phenomena" (for more discussion see
also \cite{JudgeDeskbook}.
 
In turn, to build up a dynamic model, first we derive some principles, which we call the Principles of Dynamic Modeling in Health Research. These Principles are independent of the target task, so they could be applied to any other task, for example, to follow-up  analysis with end-points.
 
The aims of this paper are as follows:
\begin{itemize}
\item to develop the Dynamic Model of Population Health for the case of one State Variable
\item to build up the Dynamic Regression Method and algorithm allowing for multiple surveys with different age ranges with large volume of data involved
\item to run the illustrative analysis
\end{itemize}
 
Note that each of the parts above is worth of more detailed, separate  presentation.
 
Therefore, the challenge was to provide concise and logically completed descriptions of all parts, clearly outlining logical interrelations between them.
 
The earlier version of the Dynamic Regression Method was developed and presented  by \cite{MolMic}, where C - trends were suggested as an alternative to circular trends, used so far.
Historically, the  method developed in MONICA for checking
consistency of the reported demographic data (\cite{Moltchanov1999})
served  as the prototype for the method developed by \cite{MolMic}.
Some general aspects, such as criteria  for commonly used health indicators to serve as  system State Variables, have been considered earlier by
\cite{Moltchanov1993}.

Section \ref{ModPopLev} describes  Dynamic Model paradigm being applied to health research on Individual and Population Level.
In Section \ref{Anaform} we
present the analytical form and general form of statistical  model for the  case of continuous, normally distributed one parameter.
The data used for analysis are described  in section  \ref{Data}.
The detailed description of the DRM "fast"  algorithm  is given in section
\ref{outlines}.
The example of application, is given in section \ref{Exa}.
Section \ref{conclu} contains  conclusion and discussion.

\section{Dynamic Model of Population\label{ModPopLev}}
\subsection {Dynamic model on individual level, notation, definitions }
To describe population history, it is convenient to use $y$-$a$ plane, where  $y$  is real-valued calendar time in years, vertical axis,   $a$ is real-valued age in years, horizontal axis, with points  specified as $(y,a)$  (vertical coordinate first). This is matrix standard for indexing elements, which will be of use later.
 
Consider a population defined on observational frame (real compact) $\mathcal{C}$:
\begin{equation}\label{eq:compact}
\mathcal{C}=\{(y,a):y \in [y_{min},y_{max}], a \in
[a_{min},a_{max}]\},
\end{equation}
 
The usual assumptions must be fulfilled for the population: it should include all the residents of a geographically outlined area with well defined administrative boundaries. And neither administrative boundaries, nor rules for residentship are changed within observational frame  $\mathcal{C}$.
 
To introduce dynamic model terminology, we consider health history of one  subject of the population. Let ${x_W}(y,a)$
be a weight of this subject at point $(y,a)$. After time period ${\Delta}t$, measured in years,  the  subject will move to the point  $(y+{\Delta}t ,a+{\Delta}t)$, located on birth cohort line for this subject. For the sake of simplicity, we assume that weight changes linearly over this time period and both points belong to the observational frame  $\mathcal{C}$. Under these assumptions, the rate of change of weight at point $(y,a)$  will be expressed as
\begin{equation}\label{eq:wevu2}
u_W(y,a) =\frac{x_W(y+{\Delta}t,a+{\Delta}t)- x_W(y,a)}{{\Delta}t }
\end{equation}
 
According to the current state of knowledge,  weight change in adult is, in fact, change in amount of body fat, which is determined by balance of calories taken with meal and burned throughout the  body  activity over a certain time period (see, for example, http://www.weightlossforall.com/calories-per-pound.htm
"One pound of body fat equals roughly 3500 calories."). Or one kilogram of body fat  equals 7716.2 calories. Using this ratio, we can calculate how balance of energy results  in weight change.
 
Let function $f_B(y,a)$ be the balance of energy in kilocalories  per time unit. Then the above verbal statement could be expressed as
 
\begin{equation}\label{eq:ba2u}
u_W(y,a)=k_{conv}\cdot f_B(y,a) ,
\textrm {where } k_{conv}=0.1296
\end{equation}
or, introducing the weight-specific Driving Force $f_W(y,a)$,
 
\begin{equation}\label{eq:we2u}
u_W(y,a)=f_W(y,a), \textrm { where } f_W(y,a)=k_{conv}\cdot f_B(y,a).
\end{equation}
 
Note that alternatively we could  define $f_W(y,a)= f_B(y,a)$, and get the equation
$u_W(y,a)=k_{conv}\cdot f_W(y,a)$, which is the form for general case (any State Variable, hence  subscript is omitted):
\begin{equation}\label{eq:we2ug}
u(y,a)=k \cdot f(y,a),   k \textrm { is a fixed coefficient.}
\end{equation}
 

Let $t$  be a relative time, $t=0$ when subject is at point  $(y,a)$. In general case, to represent time change  of  a state variable,  we use function $v(t)$ - a smoothed version of  $x(y+t,a+t)$, being free of daily cycles, which are common property of   biological indicators ( for details see  \cite{Moltchanov2012}. Then (\ref {eq:we2ug})  will be transformed into:
 
\begin{equation}\label{eq:wevu4}
u(y+t,a+t) =k \cdot f(y+t,a+t),
\textrm {where } u(y+t,a+t) =
\frac{\ud v(t)}{\ud t}
\end{equation}
 
Now we rephrase  the above description of the case using terminology introduced in \cite{Moltchanov2012} for dynamic modeling of population health:
 
An \textbf{Object} ( our  subject) is moving over time along cohort line carrying, as a system, its \textbf{State Variable} $x_W(y,a)$ (weight in our case) and being all the time affected by weight-specific \textbf{Driving Force}  $f_W(y,a)$. The \textbf{Law of Motion} postulates that change of State Variable occurs due to Driving Force at rate proportional to the value of this force, which is reflected in (\ref {eq:wevu4}).
 
For the rate of change of a State Variable $u(y,a)$, while the object moves along cohort line, we will use the term "Cohort trend" or "C-trend", and, for  linguistic convenience, we will use term "Modifier" interchangeably with term "Driving Force" as  proposed in \cite{Moltchanov2012}.
 
Note, that Driving Force, as a function of time, may be a non-continuous function. While State Variable, according to (\ref {eq:wevu4}), is a continuous function of time, moreover,  it is right-differentiable one.  That property labels  health parameters, which   may serve as  State Variables. Along with weight, the measurements of  blood pressure, cholesterol, hight, as well as schooling years, are State Variables. While current smoking status, physical activity, dietary habits, along with other behavioral characteristics, are not State Variables, rather they are Modifiers.
 
Observe that continuous function of any number of State Variables is itself a State Variable. The expression (\ref{eq:wevu4}) remains valid for this variable with the resulting Modifier being a linear combination of the contributing Modifiers.
 
In our future consideration and example we will deal with such a variable, The Body Mass Index,  defined in pseudo-code notation  as
\begin{equation}\label{eq:BMI}
BMI=\frac{ weight(kg)} {height(m)^2}
\end{equation}
 
Let $x_{BMI}$ be State Variable for $BMI$. $H$ - height  of the subject in meters, which we assume to be constant for a subject in analysis frame $\mathcal{C}$.   To get equation for Law of Motion for $BMI$, we have to divide both sides of (\ref{eq:wevu4}) by $H^2$:
 
\begin{eqnarray}\label{eq:BMIvu1}
\nonumber
u_{BMI}(y,a))&=&f_{BMI}(y,a),   \\
\textrm {where } u_{BMI}(y,a)&=&\frac{u_W(y,a)}{H^2},\\ \nonumber
f_{BMI}(y,a)=\frac{f_W(y,a)}{H^2}&=&k_{conv} \cdot
\frac{ f_B(y,a) }{H^2}=k_{conv} \cdot f_{BaH}(y,a).
\end{eqnarray}
Here we have introduced notation $f_{BaH}(y,a)$ for the Energy Balance adjusted for height.

 
\subsection {Dynamic Model on Population Level: Axiomatic Setup}
We use the term "Population Level" rather than "Population",   since,  our target population, in fact, might be  a gender-specific subpopulation.
 
Each subject may enter this population  due to birth (if $a_{min}=0$), or crossing left-low boundaries, or migration in.  Each subject may leave this population due to death, migration out or crossing the right-upper boundaries. If a subject with coordinates  $(y_0,a_0)$ is within the population during time $t$ ,  at that time it has coordinates $(y_0+t,a_0+t)$. Thus, we may say that it is moving along cohort line.

Consider all subjects having coordinates on half-open interval $( (y_0,a_0-{\Delta}a), (y_0, a_0)]$ at time  $t=0$. At time  ${\Delta}t$ all those left in population will arrive at   $((y_0+{\Delta}t, a_0-{\Delta}a +{\Delta}t ), (y_0+{\Delta}t-{\Delta}a,a_0+{\Delta}t]$.  In other words, the birth cohort of width ${\Delta}a$  moves from $(y_0,a_0)$ to $(y_0+{\Delta}t,a_0+{\Delta}t)$. We may think of such a cohort as of a container moving on plane $(y,a)$. The contents of each container in process of movement is changed due to migration and death. If the rate of contents update is negligible (say, less than  1\% per year), we may ignore it in our analysis. If not, the analysis has to take this into account.
 
Each container fits the definition of the dynamic model object, if we regard the corresponding State Variable as  mean of State Variables for currently available subjects. The dynamic equation then could be obtained from ones for each subject, having form (\ref{eq:wevu4}), by taking means of both sides:
 
Since the whole selected observational frame could be covered by collection of non-overlapping cohorts of selected width, we may conclude that, in case of population, the overall dynamic model is a collection of the dynamic  models specific for each cohort.
 
In fact, it is more convenient and straightforward to go ahead with the theoretical abstraction for birth cohort, which deals with cohort containers of
infinitesimal  age range, characterized by multidimensional
distribution of  the parameters of interest, not by physical
subjects.
 
In general case, we suggest that
there potentially exists a set of random variables (r.v.)  $X_i$,
$i=1...k$ representing the corresponding set of measurable
indicators of interest (State Variables) defined at each point $(y,a)$ of compact
$\mathcal{C}$.
 
In this paper we restrict ourselves to the case of
one indicator, so that subscript of $X$ will be omitted. To make the following description more illustrative let us keep in mind the Body Mass Index (BMI) as an example of the indicator in question.
 
We introduce the following notation:
$$
v(y,a)\doteq E(X(y,a)).
$$
 
For the sake of simplicity while describing the core dynamic model,
we assume:
 
\begin{equation} \label{eq:L301}
X(y,a)=v(y,a)+\epsilon \rm{, where }~E(\epsilon)=0,  ~ D(\epsilon)=
\sigma^2,~ \forall (y,a): (y,a)\in \mathcal{C}
\end{equation}

The dynamic equations describe changes of the distribution  of r.v.
$X$ for a birth cohort  taken at  point $(y,a)$ over time interval
$\ud t$:
 
We introduce function $u(y,a)$, rate of change of the State Variable along cohort line:
\begin{equation}\label{eq:L302A}
u(y,a) \doteq  \lim_{\Delta t\to 0}\frac {v(y+\Delta t,a+ \Delta t)-v(y,a)}{\Delta t}
\end{equation}
 
The Low of Motion postulates existing at each point $(y,a)$  such a Driving Force $f(y,a)$, which causes change of the State Variable, while point (container) moves  along cohort line, at rate proportional to the value of the force, $u(y,a) \propto f(y,a)$, or, with
$f(y,a)$ properly scaled,
\begin{equation}\label{eq:lomopo}
u(y,a) = f(y,a)
\end{equation}
 
Note that the above formulation of Low of Motion and equation (\ref{eq:lomopo}) are similar to those for model on individual level (equation (\ref{eq:we2ug}) ). The principal difference is that we are dealing now with population means $v(y,a)$ ) and its cohort trends $u(y,a)$, and population means of Modifier $f(y,a)$.
 
In this paper we consider the case, when Driving Force $f(y,a)$ does not depend  on the properties of the cohort (value of $v(y,a)$ ).
The generalization of the model  for
the case of multidimensional distribution and state-dependent
dynamics is formulated in \cite{Moltchanov2012}
 
For the sake of convenience we will use terms "Mean levels" or
"levels" for the values of function $v(y,a)$.
 
Let $v_0(y,a)$ be the value of $v(y,a)$ at low-left boundary of the
compact $\mathcal{C}$ for a (birth) cohort crossing the point
$(y,a)$:
\begin{equation}\label{eq:L303}
v_0(y,a) = v(y-\delta, a-\delta), \qquad where ~
\delta=min(y-y_{min},a-a_{min})
\end{equation}
 
Then $v(y,a)$ can be expressed as
\begin{equation}\label{eq:L304}
v(y,a)=v_0(y,a) + \int_0^\delta u(y-t,a-t)\ud t
\end{equation}

Thus, if the values of  $v_0(y,a)$ at low-left boundary and $u(y,a)$
on $\mathcal{C}$ are known, then the function $v(y,a)$ could be
evaluated for each point on $\mathcal{C}$.  In other words, we can set up  initial levels $v_0(y,a)$  and the Driving Force pattern on $\mathcal{C}$ and we can  simulate behavior of the population health in terms of mean levels  $v(y,a)$, using
(\ref {eq:lomopo}), (\ref{eq:L303}), (\ref {eq:L304}).

Observe that function $v(y+t,a+t)$ is right-differentiable on $t$, since it could be expressed as
\begin{equation}\label{eq:L305}
v(y+t,a+t)=v(y,a) + \int_0^t  f(y+\tau,a+\tau) \ud {\tau}
\end{equation}
At the same time, continuity on $y$ and $a$ is not an obligatory property of this function.

\section{Statistical Model: The core formulation \label{Anaform}}
 
\subsection {General Formulation of the Task }
 
Suppose that a set  of measurements is available  $(x_k,y_k,a_k)$,
$k=1,\ldots,K$, for  subjects of gender-specific subpopulation (men, for certainty),   selected in a set of the independent
cross-sectional surveys. We assume that for each survey, the
stratified by gender and age group random sample scheme was used.
The age group stratification could be different in different
surveys, as well as age range.
 
To start with analysis, first, we have to define the observational frame for the analysis, by setting up parameters  in (\ref{eq:compact}):
 
\begin{eqnarray}\label{eq:defcomp}
y_{min}\doteq  floor(  \min (y_k)),\; x_{min} \doteq floor(  \min (x_k)),  \\
y_{max} \doteq ceil(  \max (y_k),\; x_{max} \doteq ceil (  \max (x_k)),\; k=1,\ldots,K \nonumber
\end{eqnarray}
 
The general formulation of the task is to estimate the functions
$v_0(y,a)$ and $u(y,a)$ on $\mathcal{C}$, using the available
measurements $(x_k,y_k,a_k)$, $k=1,\ldots,K$.

To solve this problem, one option would be to formulate the
optimization problem in functional space:  to minimize  the
functional $I$:
 
\begin{equation}\label{eq:L305}
I(u,v_0) = \Bigg ( \sum \Big( x_k-v(y_k,a_k) \Big ) \Bigg )^2,
\end{equation}
 
applying some additional requirements on functions  $u(.,.)$ and
$v_0(.,.)$, such as   continuity (piece-wise continuity), and/or
restricted variation.

However, it seems more convenient  to transform the above problem
into the discrete - scale  analogue and to take the advantage of the
simplicity of  the analysis and  adaptation of the numerical methods
available in the standard statistical packages.
 
\subsection {Discrete-Scale Model}
 
Let $i$ and  $j$  be an integer value of  time  in years and an
integer value of age in years correspondingly. Our intention is to
build up the integer-values  proxies of the equations
($\ref{eq:L301}$ - $\ref{eq:L305}$).
 
Let $P(i,j)$  be a parallelogram-shaped element (convex hull)
defined by its angle points: $$  \{(i,j-1),~ (i,j),~ (i+1,j+1),~
(i+1,j)\}
$$ excluding its left and upper boundaries, which could be written
as
\begin{equation} \label{eq:B005}
P(i,j) \doteq \{(a,y): y \in [i,i+1),~ a \in ((j-1) +
(y-i),~j+(y-i)]\}
\end{equation}
 
We impose for function $u(.,.)$ the conditions of being constant on
each $P(i,j)$ and for functions  $v(.,.)$  -  being constant on $a$ and
linear on $y$ with constant  slope $u(i,j)$.
 
Formally this could be expressed as follows:
 
\begin{equation} \label{eq:B010}
u(y,a)=u(i,j), ~ \forall  i,j,y,a:~ (y,a) \in P(i,j)
\end{equation}
 
\begin{equation} \label{eq:B020}
v(y,a)=u(i,j) \cdot (y-i)+ v(i,j), ~\forall  i,j,y,a:~ (y,a) \in
P(i,j)
\end{equation}

We derive minimal and maximal values for $i$ and $j$ from the
correspondent values for $y$ and $a$ using definition
($\ref{eq:B005}$):
\begin{eqnarray}\label{eq:B301}
(i_{min},j_{min}): (y_{min},a_{min}) \in  P(i_{min},j_{min}), \\
(i_{max},j_{max}): (y_{max},a_{max}) \in
P(i_{max},j_{max})\nonumber
\end{eqnarray}
 
For convenience,  from now on we will use relative scale for age and
time, defined by transformation
$$
i - i_{min} \rightarrow i\rm{, }~ j - j_{min} \rightarrow j
$$

Consider functions $u(i,j)$ and $v(i,j)$ defined on integer-valued
two-dimensional domains
\begin{eqnarray}\label{eq:uvdef}
\mathcal{U}&=&\{(i,j):i \in [0,I],~ j \in [0,J]\}, \nonumber\\
\mathcal{V} &=& \{(i,j):i \in [0,I+1],~j \in [0,J+1]\},
\end{eqnarray}
 
correspondingly, where
$$
I=i_{max} -i_{min}\rm{, }~ J=j_{max} -j_{min}
$$
 
Now the main dynamic equation ($\ref{eq:L302A} $ ) could be rewritten
as
 
\begin{equation}\label{eq:B302}
v(i+1,j+1)=v(i,j)+u(i,j),  ~\forall (i,j)\in  \mathcal{U}
\end{equation}
 
Let $v_0(i,j)$ be the value of $v(.,.)$ at low-left boundary of the
domain $\mathcal{V}$ corresponding to a (birth) cohort crossing the
point $(i,j)$:
\begin{equation}\label{eq:B303}
v_0(i,j) = v(i-\delta, j-\delta), \rm{where }¨~  \delta=min(i,j).
\end{equation}
 
Combining ($\ref{eq:B302}$) and  ($\ref{eq:B303}$), we rewrite
equation (\ref{eq:L303})  as:
 
\begin{equation}\label{eq:B304}
v(i,j) =v_0(i,j) + \sum_{m=1}^\delta u(i-m,j-m)
\end{equation}
 
From ($\ref{eq:B304}$), it follows that if $v(i,j)$ is set up on the
low-left boundary of  $\mathcal{V}$ and $u(i,j)$ is set up on the
whole $\mathcal{U}$ then $v(i,j)$ could be calculated for the whole
$\mathcal{V}$.
 
Finally, assembling ($\ref{eq:L301}$), $\ref{eq:B304}$) and
($\ref{eq:B020}$) for each available observation  $(x_k,y_k,a_k)$,
$k=1,\ldots,K$, we obtain:

\begin{eqnarray}\label{eq:B305}
x_k =v_0(i,j) + \sum_{m=1}^\delta u(i-m,j-m) + (y_k-i)\cdot u(i,j)
+\epsilon_k \rm{,}
{}\nonumber\\
\rm{ where }~Var(\epsilon_k)=\sigma^2, ~
Cov(\epsilon_k,\epsilon_l)=0,~if~ k \neq l
\end{eqnarray}

Let $\mathbf{z}$  be a vector  with components $v_0(i,j)$ and
$u(i,j)$ ordered in the following way:
 
\begin{eqnarray}\label{eq:vuz}
\mathbf{v}_0 &=& \big(v(I+1,0),\ldots,v(0,0), \ldots,
v(0,J+1)\big)^T \nonumber\\
\mathbf{u} &=& \big(u(0,0),\ldots,u(0,J),\ldots,
u(I,0),\ldots,u(I,J) \big)^T \nonumber\\
\mathbf{z} &=& \big( \mathbf{v}_0^T \quad \vline \quad
\mathbf{u}^T\big)^T
\end{eqnarray}

Using vector $\mathbf{z}$  and introducing vector of coefficients
$\mathbf{b}_k$, we  can rewrite ($\ref{eq:B305}$) in the form
\begin{equation}\label{eq:B306}
x_k =(\mathbf{b}_k, \mathbf{z}) +\epsilon_k, ~\rm{where }
~Var(\epsilon_k)=\sigma^2, ~ Cov(\epsilon_k,\epsilon_l)=0,~if~ k
\neq l
\end{equation}

This form represents a particular case of Gauss-Markov Setup for the
Least Squares Linear Estimation problem \cite{Rao}.
 
Let $\mathbf{B_0}$ be a matrix composed of row vectors
$\mathbf{b}_k^T $ in  ($\ref{eq:B306}$), $\mathbf{z} $ and
$\mathbf{x}_0 $ stand for column vectors of the parameters
$\mathnormal{z}_j $ and the variables $\mathnormal{x}_k $
correspondingly, and $\mathnormal{S}_0 $  be a scalar function
defined as
\begin{equation*}\label{eq:B307}
\mathit{S}_0( \mathbf{z} ) =( \mathbf{B_0}\mathbf{z} -
\mathbf{x}_0)^T  ( \mathbf{B_0}\mathbf{z} -  \mathbf{x}_0)
\end{equation*}
 
Note that if $ \mathit{rank}(\mathbf{B_0} ) = \mathit{dim}(
\mathbf{z} ) $, then  estimates obtained by unconditional minimizing
of function $\mathnormal{S}_0 ( \mathbf{z} ) $  are unique ones.
Such a case takes place only if the observations cover all the
elements   $P(i,j)$  when  surveys cover the whole analysis period
without gaps.
 
In practical cases, minimizing of  $\mathnormal{S}_0  $ results in
singular or ill-posed Inverse Problem, and  so-called regularization
techniques  are needed to obtain meaningful solution estimates. Most
of these techniques employ the idea of smoothing of some function
having clear physical interpretation \cite{Neumaier}.
 
Here we suggest several components for smoothing functions $v(.,.)$  and $u(.,.)$.
\subsection{Smoothing}
 
We define the following indicator of smoothness of function  $u(.,.)$ (C-trends);
\begin{eqnarray*}
\mathnormal{S}_1( \mathbf{z} ) & =  \sum_{i=0}^{I} \sum_{j=1}^{J-1}
\Big( u(i,j-1)-2u(i,j)+u(i,j+1) \Big)^2
+ {}\nonumber\\
&  \sum_{j=0}^{J} \sum_{i=1}^{I-1}    \Big( u(i-1,j)-2u(i,j)+u(i+1,j)  \Big)^2
\end{eqnarray*}
allowing form
 
\begin{equation}\label{eq:S1}
\mathit{S}_1( \mathbf{z} ) =( \mathbf{B_1}\mathbf{z}- 0 )^T  (
\mathbf{B_1}\mathbf{z} - 0 )
\end{equation}

We define the following  indicator of smoothness of function
$v(.,.)$
\begin{eqnarray}\label{eq:B401}
\mathnormal{S}_2( \mathbf{z} ) & =  \sum_{i=0}^{I+1} \sum_{j=1}^{J}
\Big( v(i,j-1)-2v(i,j)+v(i,j+1) \Big)^2
+ {}\nonumber\\
&  \sum_{j=0}^{J+1} \sum_{i=1}^{I}    \Big( v(i-1,j)-2v(i,j)+v(i+1,j)  \Big)^2
\end{eqnarray}
Each term in this sum represents the  square for a proxy  of the
second derivative of function $v(.,.)$  with respect to age or with
respect to calendar time at point $(i,j)$.
 
Replacing $v(.,.)$ by $v_0(.,.)$ and $u(.,.)$ using
($\ref{eq:B304}$), and the last ones by vector  $\mathbf{z}$, we
will transform the previous expression to the following form:
\begin{equation}\label{eq:B402}
\mathit{S}_2( \mathbf{z} ) =( \mathbf{B_2}\mathbf{z}- 0 )^T  (
\mathbf{B_2}\mathbf{z} - 0 )
\end{equation}

Now we can add one or both constraints  ${\mathit{S}_k( \mathbf{z} )
\leq \alpha_k }$ with some selected  ${\alpha_k \geq 0}$,  $k=1,2$,
to the model (${\ref{eq:B306}}$). Observe that indicators
$\mathit{S}_0$, $\mathit{S}_1$, $\mathit{S}_2$ are quadratic
functions in finite vector space $ \mathrm{E}_n $ with elements
(vectors) $\mathbf{z}$ and $n=dim( \mathbf{z} )$. The optimization
problem for point estimation for our case,  could be formulated as
\begin{equation}\label{eq:P1}
\min_{\mathbf{x} \in \rm{E}_n }  S_0( \mathbf{x} ) \textrm{, subject
to } {S_k( \mathbf{x} ) \leq \alpha_k } \textrm{, with given
}{\alpha_k
>0 }, ~k=1,2.
\end{equation}
Let $n_0$ , $n_1$ and $n_2$  be numbers of rows in matrices
$\mathbf{B}_0$ , $\mathbf{B}_1$ and $\mathbf{B}_2$ correspondingly. Let
$\lambda_1$,  $\lambda_2$  be some non-negative  scalars.
Introducing matrices and vectors
\begin{equation}\label{eq:p11}
\mathbf{B}= \left( \begin{array}{c}
\mathbf{B}_0 \\
\hline
\mathbf{B}_1 \\
\hline
\mathbf{B}_2 \\
\end{array} \right),
\quad \mathbf{x}= \left( \begin{array}{c}
\mathbf{x}_0 \\
\hline
\mathbf{0}\\
\hline
\mathbf{0}\\
\end{array} \right),
\quad \mathbf{W}= \left( \begin{array}{c|c|c}
\mathbf{I}_0 & 0 & 0 \\
\hline
0 & \lambda_1 \mathbf{I}_1 & 0\\
\hline
0 & 0 & \lambda_2 \mathbf{I}_2 \\
\end{array} \right)
\end{equation}
where $\mathbf{I}_0$, $\mathbf{I}_1$ and  $\mathbf{I}_2$ are identity matrices of rank  $n_0$ , $n_1$ and $n_2$ correspondingly, we
can formulate the problem of least squares estimation in the
following form (a modification of Gauss-Markov setup which fits form
of Aitken setup \cite{Rao}
\begin{equation}\label{eq:p12}
\mathbf{x}=\mathbf{B}\mathbf{z}+\mathbf{\epsilon},\quad
E(\mathbf{\epsilon})=\mathbf{0}, \quad
D(\mathbf{\epsilon})=\sigma^2\mathbf{W}^{-1}
\end{equation}
for which the point estimation problem is
 
\begin{eqnarray}\label{eq:p13}
&\min_{\mathbf{z} \in \mathrm{E}_n }  S( \mathbf{z}), \\
\nonumber
&\textrm{ where } S( \mathbf{z} ) =( \mathbf{B}\mathbf{z} - \mathbf{x})^T
\mathbf{W} (\mathbf{B}\mathbf{z} -  \mathbf{x})= S_0
(\mathbf{z})+\lambda_1 S_1(\mathbf{z})+\lambda_2 S_2(\mathbf{z})
\end{eqnarray}
 
In \cite {MolMic} it was shown that problems (\ref{eq:P1}) and (\ref{eq:p13}) are equivalent:  problem  (\ref{eq:P1}) with given
$\alpha_1$, $\alpha_2$  possesses  the same solution as problem  (\ref{eq:p13}) with some $\lambda_1$, $\lambda_2$, and vice versa, or both don't possess any solution.
In particular, it was shown that for  existence of a unique  solution to problem (\ref{eq:p13}) it
is sufficient to have 4 data points such that the corresponding
points $(y,a)$ on plane $y,a$ satisfy condition: no any 3 of
them are located on a common straight line.
 
More discussion on statistical properties of problem (\ref{eq:p13}) could be found in  \cite{Moltchanov2012}.
 
As soon as parameters $\lambda_1$,  $\lambda_2$  are given in
setup (\ref{eq:p11}, \ref{eq:p12}), the following could be obtained routinely:
$\mathbf{\hat{z}}$  - point estimate of vector $\mathbf{z}$,
covariance matrix of this estimate $Cov(\hat{\mathbf{z} })$, and $\hat{\sigma^2}$ - estimate of ${\sigma^2}$.
 
The task remains, to formulate criteria of smoothness of a solution above  and to organize an iterative process  by selecting   $\lambda_1$, $\lambda_2$ and solving Linear Regression Problem   (\ref{eq:p12}) at each iteration, until the predefined criteria  are satisfied.
 
The design of the corresponding algorithm should reflect substantially the size and the structure of data to be analyzed.

\section{Data\label{Data}}
To illustrate the method and to demonstrate its performance, the
data will be used  comprising 7 cross-sectional surveys, conducted in North Karelia,
Finland,  during the period  1972 -2002. Details of these data are shown in
Table \ref{tbl:data}:
\begin{itemize}
\item Study population:  North Karelia, Finland, men.
\item  Study period: 1972-2002.
\item  Source of data: cross sectional independent surveys conducted in
years  1972 -2002 every 5 years.
\item Sampling frame: simple random sample scheme was used  in years 1972, 1977, in other years the stratified by 10-year age groups
(25-34, 35-44, 45-54, 55-64 (65-74 if available)) (and gender) random sample scheme was used.
\item in year 1972 survey was conducted in the period of 8 months, February - September, in year 1997 - period was 6 months, January-June,
in  other years surveys were conducted in 4 or 3 month, starting in January.
\item  Participation rate  varies from 66\%  to 94\%
\item Total number of observation with non-missing  BMI,  age and date of examination, is 11045.
\item Age ranges are different in different surveys: 25-59 for year 1972, 25-64 for years 1977-1992, 25-74 for years 1997, 2002.
\end{itemize}

Original measurements of interest were gender, date of
birth, date of examination, weight and  height. At the study sites, height and weight were measured  using a standardized protocol. Height was measured to the nearest 0.1 cm. Body weight of the participants wearing usual light indoor clothing without shoes was measured with a 0.1 kg precision on a balanced beam scale.
 
The analysis variables included in the model are:
 
BMI - the Body Mass Index, defined as $ {weight (kg)} / {height
(m)}^{2} $.
 
AGE - age in full years, defined as  year of examination  minus year
of birth.
 
YEAR - date of examination measured in years.

\section{Outlines of the Algorithms for large scale data. Smoothing  criteria \label{outlines}}
 
\subsection{Data aggregation}

The original individual data might be of quite large  size (11045 in our example), which equals to rows number $n_0$ of matrix $\mathbf{B}_0$ in (\ref{eq:p11}),  and, along with columns number, proportional to product of year range by  age range,  this results in resource-expensive computation at each step of iteration.
 
Data aggregation reduces significantly number of rows $n_0$. Aggregation is applied to the original measurements  $(x_k,y_k,a_k)$, $k=1,\ldots,K$, producing summary statistics for $(age \cdot  year)$ cells with size 1, containing arithmetic means   $(\overline{x}_c , \overline{y}_c , \overline{a}_c  )$, and number of original measurements  in each cell  $(n_c)$, where $c=1,\ldots,C$ - collection of non-empty cells. In general case, the cells are excluded with
$n \leq n_{exc}$.  In our example, $n_{exc}$ is set to 5, and there were no excluded cells, and  number of cells is equal to sum of all age ranges.
 
For convenience, we rename the aggregated  data to the form of the original ones, $(x_k,y_k,a_k)$, $k=1,\ldots,K$, where $K=C$. After that all the above formulas  remain valid.

\subsection{Analysis Domain}
Each data point at cell $(i,j)$ causes inclusion into the analysis $u(i,j)$ from  the cells $(i-k,j-k)$, $k=0,... min(i,j)$. Overlapping of such cells creates collection of cohort segments starting at some points on low-left border of domain $\mathcal{U}$. Some of such segments contain several data points, other only one.
 
For future use, we define selection options:
\begin{itemize}
\item   Domain=1: all cohort segments are selected;
\item   Domain=2: only cohort segments with 2 and more data points are selected.
\end{itemize}
 
At the next step, selected collection  is modified by inclusion additionally those cells which  fill the internal gaps in all vertical and horizontal segments. The resulting  analysis domain   $\mathcal{U}_{a}$, may contain significantly  less cells, compared with original  $\mathcal{U}$, thus, diminishing number of columns in the analysis problem. We introduce vector $\mathbf{u}_{ind}$, indicating (with 1/0) the components of vector  $\mathbf{u}$, corresponding to domain   $\mathcal{U}_{a}$ and vector $\mathbf{u}_a$ part of vector  $\mathbf{u}$  with components included in the analyses.
 
Domain   $\mathcal{U}_{a}$ also determines subset of components of vector  $\mathbf{v_0}$ included in the analysis, which is a segment described by $i_l$ and $i_r$ indexes of the first and last included components.

\subsection{Between-cohorts smoothness  instead of smoothness of $v(.,.)$.}
We replace indicator of smoothness (\ref{eq:B401}) by the following one:
 
\begin{equation}\label{eq:smoco}
\mathnormal{S}_2( \mathbf{z} )  =  \sum_{i=i_l+1 }^{i_r-1}
\Big(  v_{0,i-1}-2v_{0,i}+v_{0,i+1}  \Big)^2,
\end{equation}
allowing form  (\ref{eq:B402}) with the corresponding  matrix $\mathbf{B_2}$.
 
\subsection{Iteration step analysis outcomes and control of iterations.}
Let  $\mathbf{z}_{ind}$ be index vector  indicating (with $1$) the components  of the original vector $\mathbf{z}$ (\ref{eq:B401}) included in the analysis.
 
Let, further, $\mathbf{z}_{a}$  be subset of the components of   $\mathbf{z}$ - vector of parameters to be estimated at each step of iteration. Correspondingly, all design matrices
will be modified
 
$\mathbf{B_i} \rightarrow \mathbf{B_{a,i}}$
 
by excluding the columns for components, not participating in the analysis.
 
Besides, in matrix $\mathbf{B_1}$ all the rows will be excluded, where not all the components belong to domain $\mathcal{U}_{a}$
 
After point estimates  $\mathbf{\hat{z}_{a}}$,
covariance matrix of this estimate $Cov(\hat{\mathbf{z_a} })$, and estimate of ${\sigma^2}$ will be found,   we recalculate this solution to the original scale, getting
$\mathbf{\hat{z}}$, $Cov(\hat{\mathbf{z} })$ and also $Corr(\hat{\mathbf{z} })$ - Pearson correlation  matrix.
For this purpose we will use matrix
$\mathbf{A_{za2z}}:  \mathbf{z} =\mathbf{A_{za2z}}\mathbf{z_a}$, in which  we set all rows for non-participating components to missing values(.). So, that matrices $Cov(\hat{\mathbf{z} })$ and  $Corr(\hat{\mathbf{z} })$ have non missing values only if both components are "participating".
 
Let  $r_{i,j,j+1}$ be coefficient of correlation of $u_{i,j}$ and $u_{i,j+1}$ - horizontal link, and   $r_{i,j,i+1}$ be coefficient of correlation of $u_{i,j}$ and $u_{i+1,j}$ - vertical link. Observe that the closer to 1 the coefficients of correlations are, the more "smooth" is the plot  of  values of  $u_{i+,j}$ horizontally or vertically.

We define the following measure of smoothness of surface $u(i,j)$:
 
\begin{eqnarray}\label{eq:smoru}
\bar{r}_u = \frac{1} {n_{ru}}  \Big( \sum_{i=0}^{I} \sum_{j=1}^{J-1}
r_{i,j,j+1} +
\sum_{j=0}^{J} \sum_{i=0}^{I-1}  r_{i,j,i+1}\Big) ,
\end{eqnarray}
 
where sum is taken only over non-missing values, and  $n_{ru}$ is number of such a values.
Thus, $\bar{r}_u$ is an average coefficient of correlation between two adjacent horizontal or vertical $u(i,j)$ valus in domain  $\mathcal{U}_{a}$.
 
Similarly, we define measure of smoothness for initial values of cohorts
 
\begin{eqnarray}\label{eq:smorv}
\bar{r}_v = \frac{1} {i_r-i_l +1}  \sum_{i=i_l}^{i_r-1}
r_{i,i+1}.
\end{eqnarray}

The target condition  for analysis is set up in terms of reference values $r_u$ and  $r_v$ and accuracy levels  $\delta_u$, $\delta_v$  . At the end of each iteration the following condition is checked:
 
\begin{eqnarray}\label{eq:cond}
(abs (log \Big(\frac{1-\bar{r}_u^2 } {1-{r}_u^2}\Big) )) \leq \delta_u) \; AND \;
	 (abs (log \Big(\frac{1-\bar{r}_v^2 } {1-{r}_v^2}\Big) )  \leq \delta_v).
\end{eqnarray}
If this condition is fulfilled, then iterations stop.
Otherwise, the new  values for weights are defined as follows:
 
\begin{eqnarray}\label{eq:newla}
{\lambda}_{1,new} = {\lambda}_1 \Big( \frac{1-\bar{r}_u^2 } {1-{r}_u^2} \Big), \;
{\lambda}_{2,new}={\lambda}_2 \Big(\frac{1-\bar{r}_v^2 }  {1-{r}_v^2} \Big)
\end{eqnarray}
and the next iteration is started.

To measure difference in C-trends over age and calendar year, the pairwise comparison tests are performed for
mean values of C-trends, evaluated  for a set of age-year clusters, defined  by cluster sizes, $\Delta_a$ and $\Delta_y$.
 
Let  $\mathbf{U_{c}}$ be matrix of such mean values,   $\mathbf{u_{c}}$=$Shape(\mathbf{U_{c}},1)^T$ and matrix  $\mathbf{A_{u2uc}}$ such that  $\mathbf{u_{c}}= \mathbf{A_{u2uc}} \mathbf{\hat{u}}$. As soon as, matrix $\mathbf{A_{u2uc}}$ is created for given  $\Delta_a$, $\Delta_y$, $\mathbf{U_{c}}$  could be calculated, as well as  variance/covariance values for its elements in a format of

$\mathbf{C}=  Cov(\mathbf{\hat{u_c}})= \mathbf{A_{u2uc}} Cov(\hat{\mathbf{u} }) \mathbf{A_{u2uc}}^T$.
 
For each cluster, statistics and corresponding probabilities are
computed for  pairwise comparison of mean C-trends for current cluster
and  for adjacent one for older age group, and for current one and
for adjacent one for the next calendar years period ( if the
corresponding clusters exist). Using classical paradigm, this is done by testing linear
hypotheses in form
 
$\mathbf{H}_0: {u_c}_i-{u_c}_j=0$.
 
General expression for F-value (see for,
example, SAS/Stat manual, \cite{SAS9.3STAT}) in this case takes a simple form
\begin{equation*}\label{fexpr}
F = \frac   {({u_c}_i  -   {u_c}_j)^2 } { c_{i,i}- 2 c_{i,j}+ c_{j,j}}
\end{equation*}
Corresponding probability is computed using SAS function $probF$ (see
\cite{SAS9.3L}) as
 
$ Pr = 1- probF(F,1,n-r)$
 
Note, that in Bayesian view, these probabilities should be referred to as tail-area probabilities for posterior predictive distributions (\cite {Gel1} , p.169).
 
Results of pairwise tests are presented graphically in figure, produced  by PROC GCONTOUR, properly annotated ( see Figure \ref{Chart} in example of application).
 
To asses goodness of fit, $\mathnormal{R}^2$ is calculated:
\begin{equation}\label{eq:Rsq}
\mathnormal{R}^2
= 1-\frac{(\mathbf{B_0}\mathbf{\hat{z}}- \mathbf{x_0})^T \cdot
(\mathbf{B_0}\mathbf{\hat{z}}- \mathbf{x_0}) }
{\sum_{c=1}^{C} {\big( x_{0,c}- \frac{1}{C} \sum_{i=1}^{C}{x_{0,i} }    \big)^2}}
\end{equation}

The algorithm, implementing the above outlines,  is written in SAS code using SAS products
(\cite{SAS9.3L},  \cite{SAS9.3PROC}, \cite{SAS9.3IML}, \cite{SAS9.3STAT}  \cite{SAS9.3GRAPH}).

For reference, we will call this algorithm DRM3(A), with prefix DRM3 to differentiate it from those developed in \cite{Moltchanov2012}:
DRM2(R) for "oRiginal" data, and DRM2(A) - for "Aggregated" data.

\section{Example of Application \label{Exa}}

\subsection{Analysis Setup}
\begin{figure}
\begin{center}
\centerline{\includegraphics[angle=90,width=20.5cm]{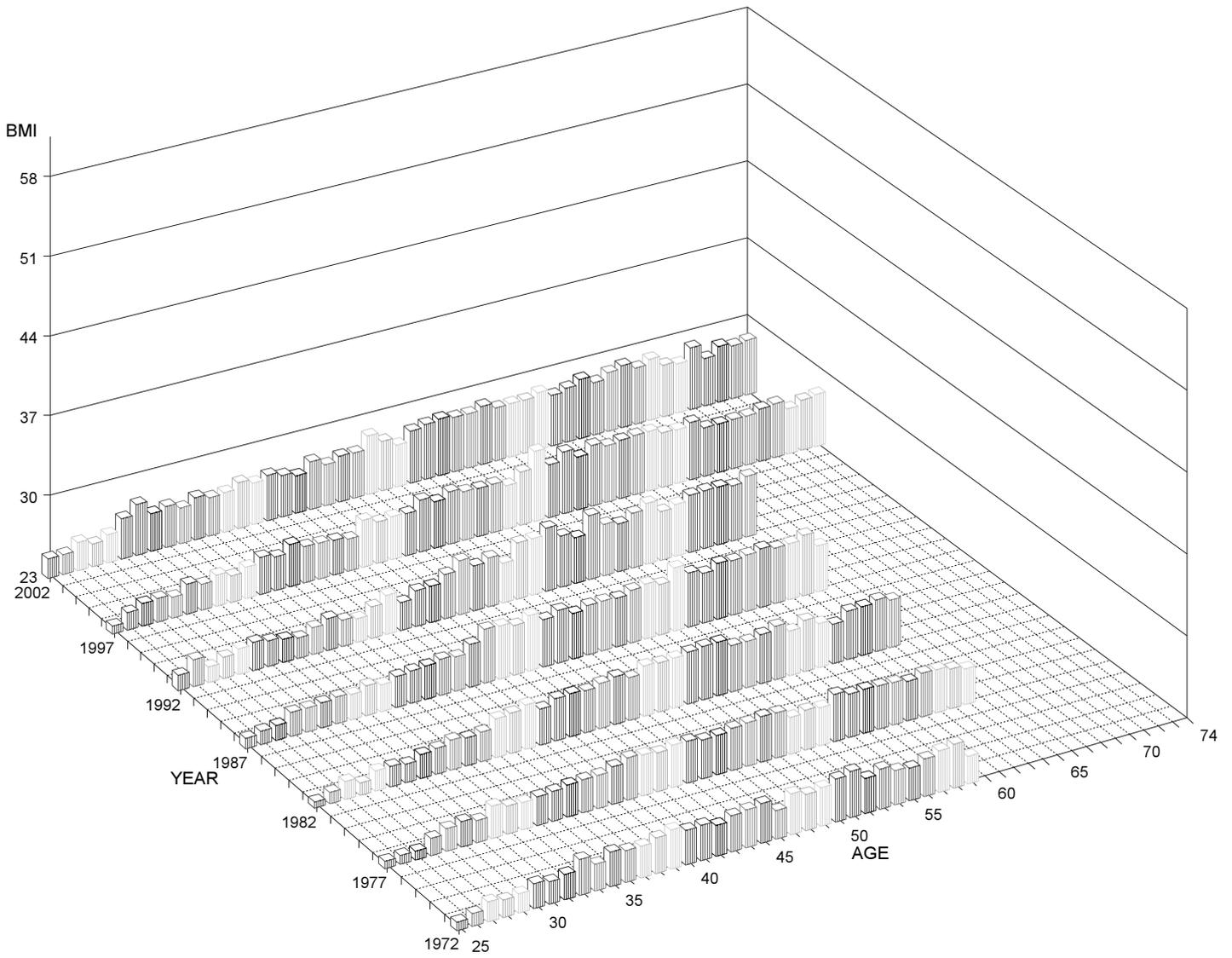}}
\end{center}
\caption{BMI, Men. Survey data. Means by year and age.
\label{ori}}
\end{figure}

The algorithm modification DRM3(A) is used, preprocessing original data, described in Table \ref{tbl:data}, into aggregated format. These data are shown in Figure \ref{ori}.
The analysis was set up for the  (maximal) age range 25-74 and for the calendar
year period 1972-2002.

To control iterations, the accuracy levels  were selected $\delta_u=0.05$, $\delta_v=0.05$. We have run analysis for several pairs of reference levels $r_v$ and $r_u$.
Let  $R(r_v,r_u)$  be set of outcomes (in terms of estimates, figures and tables) for analysis run with $r_v,r_u)$.
As a main outcome, we present analysis  $R(0.7,0.9)$.  For comparison, we have produced also analyses  $R(0.7,0.7)$, $R(0.7,0.8)$ and  $R(0.7,0.95)$.
 
\subsection{Outputs}

The results of the each analysis are visualized the set of
3-dimensional figures and special plots.
 
First, we present results for analysis  $R(0.7,0.9)$.
 
Figure~\ref{ori} displays the values representing  means of BMI
calculated for each age and year, for which the survey data are
available (number of cases in each cell exceeds 9). To visualize the
along-cohort changes, the columns corresponding to the same birth
cohorts in different surveys are drawn using similar shades of gray.
 
Note that this figure is the same for different analyses $R(r_v,r_u)$.
 
\begin{figure}
\begin{center}
\centerline{\includegraphics[angle=90,width=20.5cm]{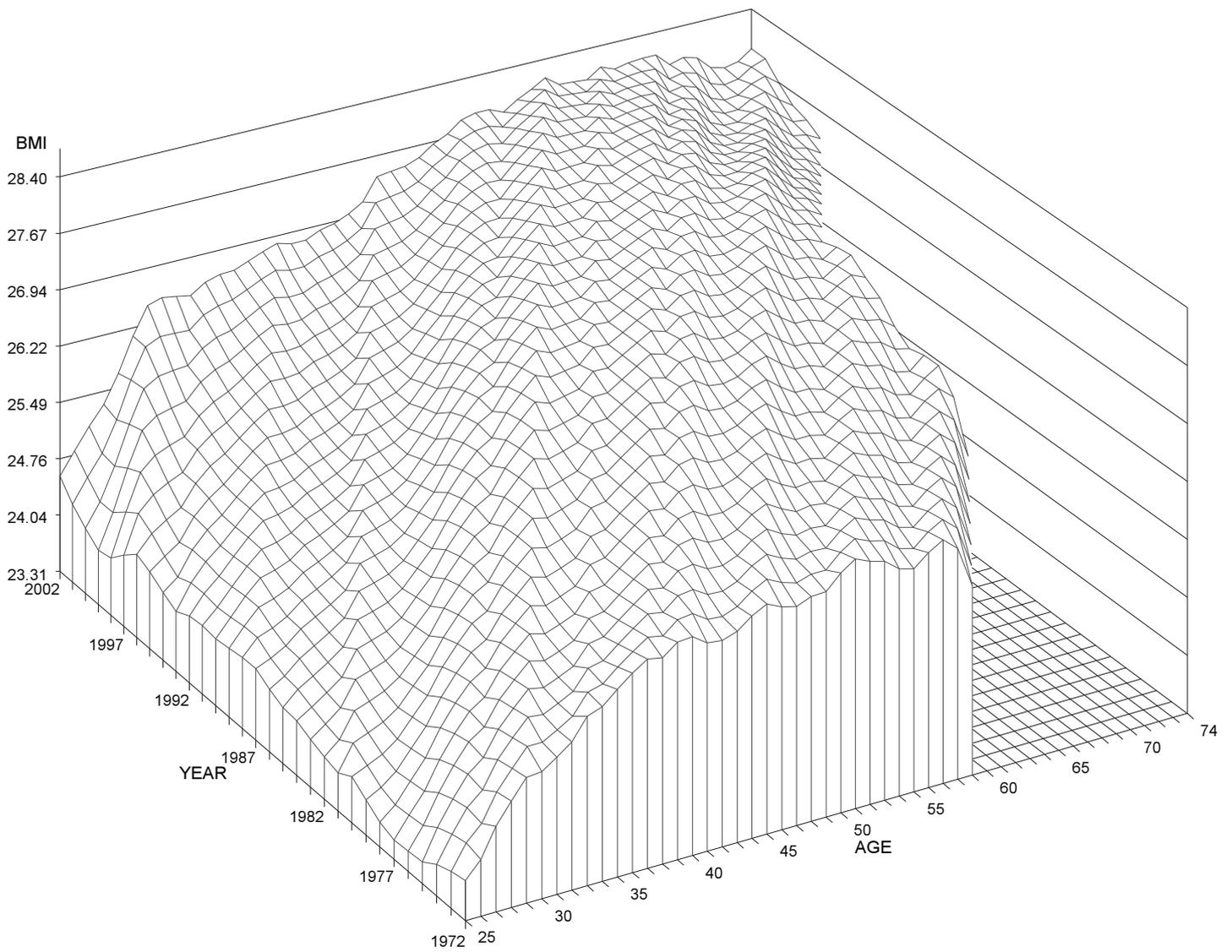}}
\end{center}
\caption{R(0.7,0.9) BMI, Men. Estimates of means by year and age.
\label{v}}
\end{figure}

Figure~\ref{v} displays estimates for the mean levels of BMI for the whole
domain, with study age range plus one year, and study period plus one
year.
 
\begin{figure}
\begin{center}
\centerline{\includegraphics[angle=90,width=20.5cm]{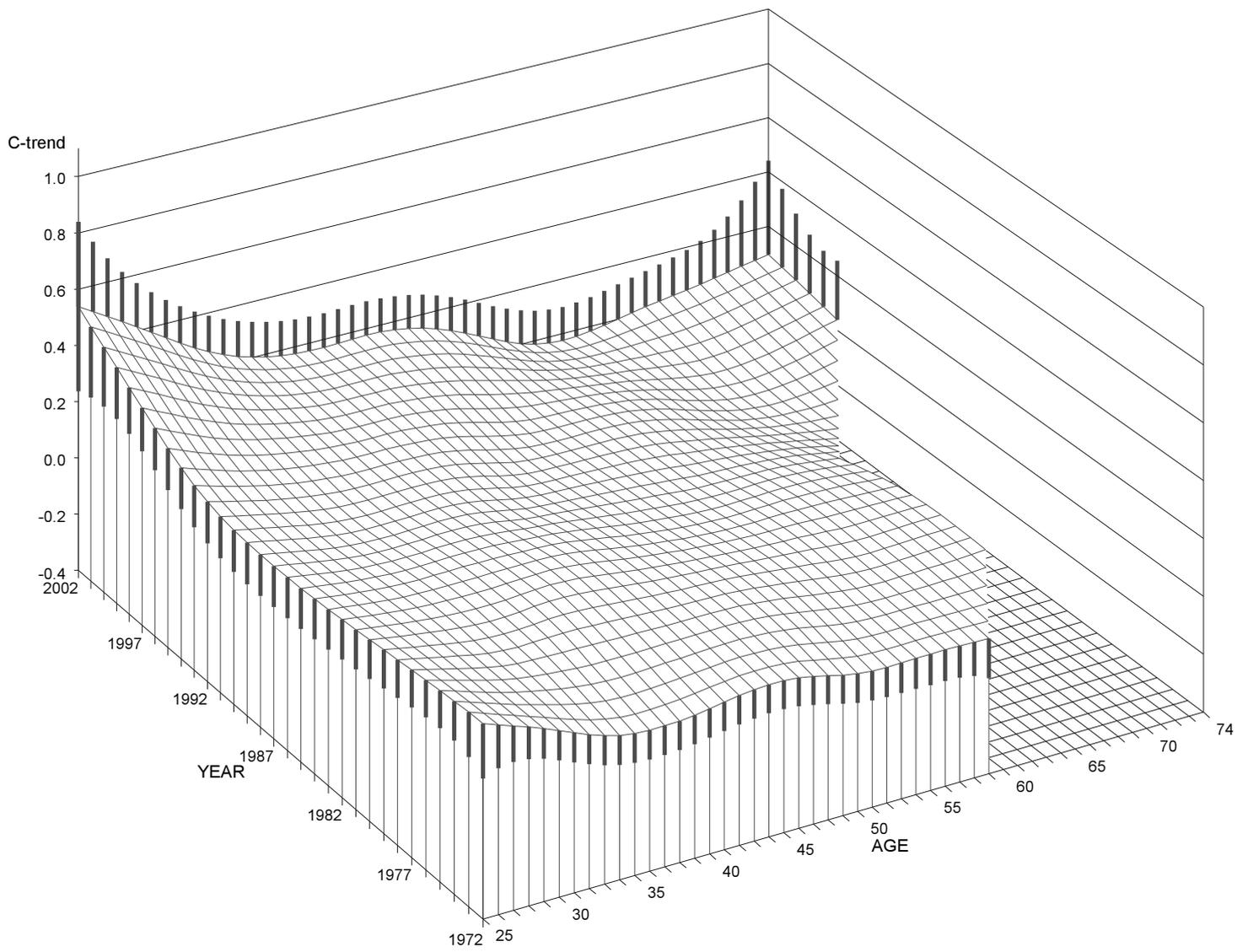}}
\end{center}
\caption{R(0.7,0.9), $\mathnormal{R^2}=0.92$.  BMI, Men. Estimates of C-trends by year and age.
\label{u}}
\end{figure}

Figure~\ref{u} displays C-trends with upper or low part of the 95\% confidence intervals, shown at boundaries only.
 
\begin{figure}
\begin{center}
\centerline{\includegraphics[angle=90,width=20.5cm]{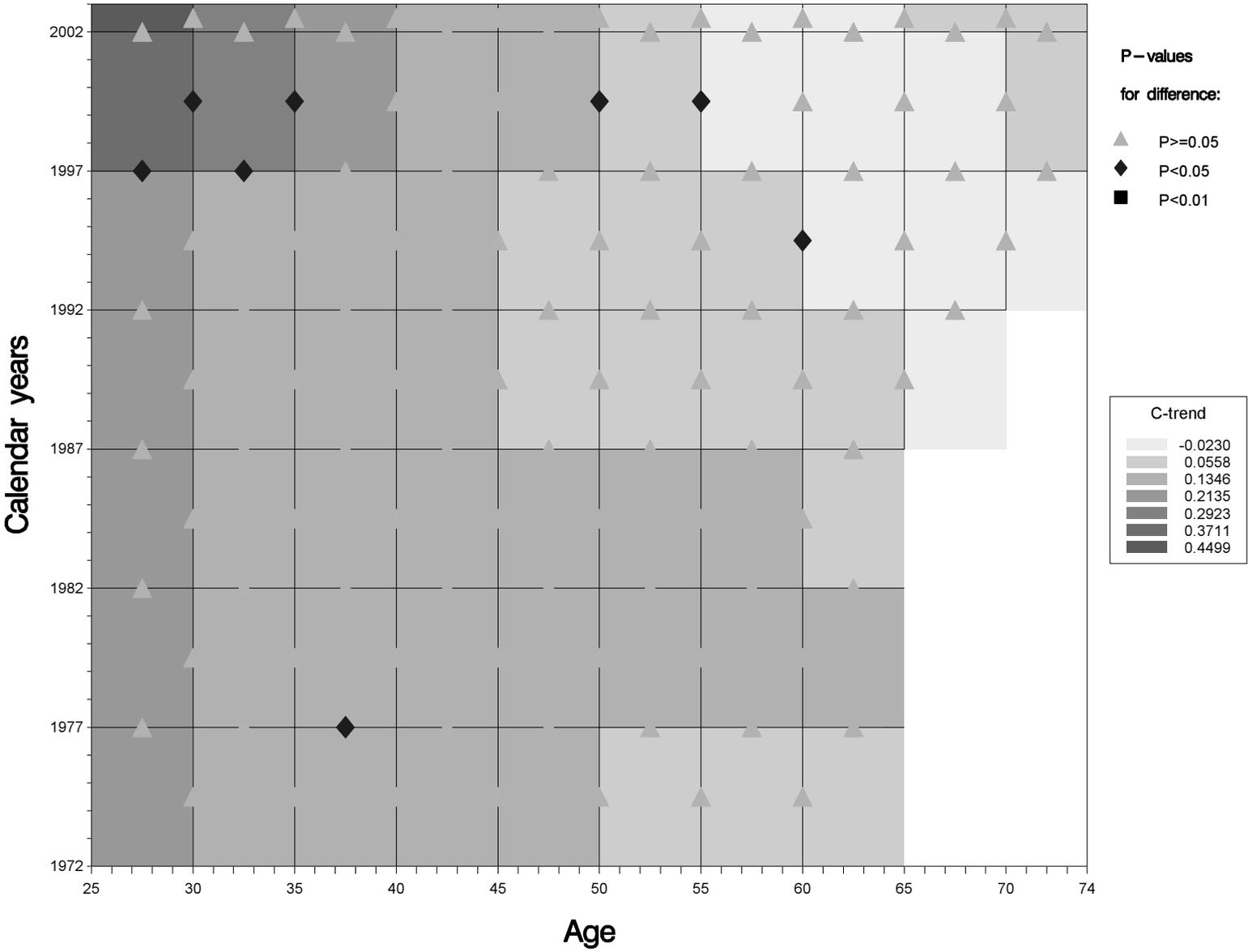}}
\end{center}
\caption{R(0.7,0.9) BMI, Men. Comparison of means of C-trends  by clusters of age and calendar years.
\label{Chart}}
\end{figure}

Figure~\ref{Chart} displays mean levels of C-trends  for specified  5-year age-year clusters, with P-values for differences between clusters.
 
\begin{figure}
\begin{center}
\centerline{\includegraphics[angle=90,width=20.5cm]{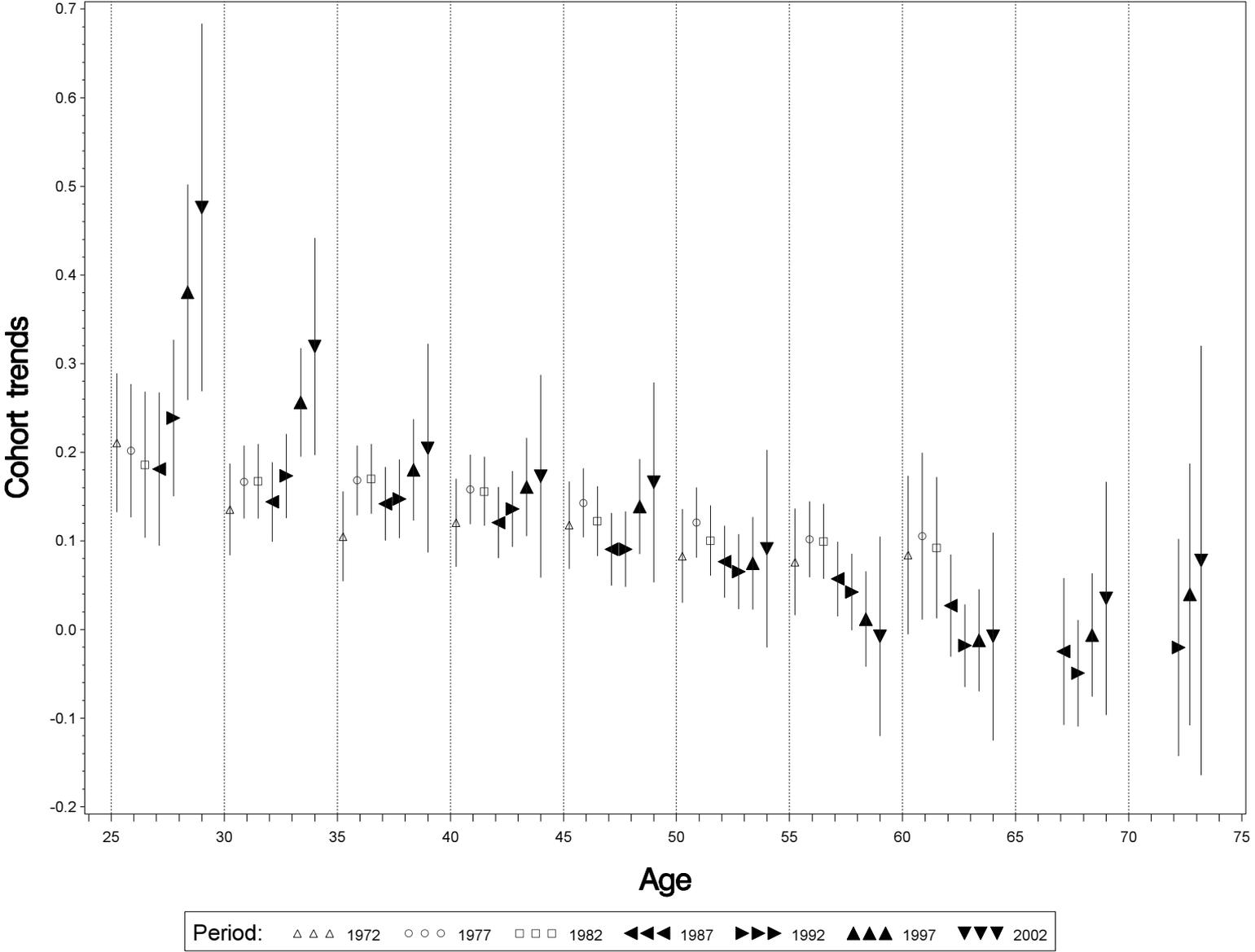}}
\end{center}
\caption{R(0.7,0.9) BMI, Men. C-trends (means and CI)  by  clusters of age and calendar years.
\label{u_CI}}
\end{figure}

Figure~\ref{u_CI} displays mean C-trend with 95\% Confidence Interval for each cluster.
For each points, horizontal position indicates age range of the  correspondent cluster.
 
\begin{figure}
\begin{center}
\centerline{\includegraphics[angle=90,width=20.5cm]{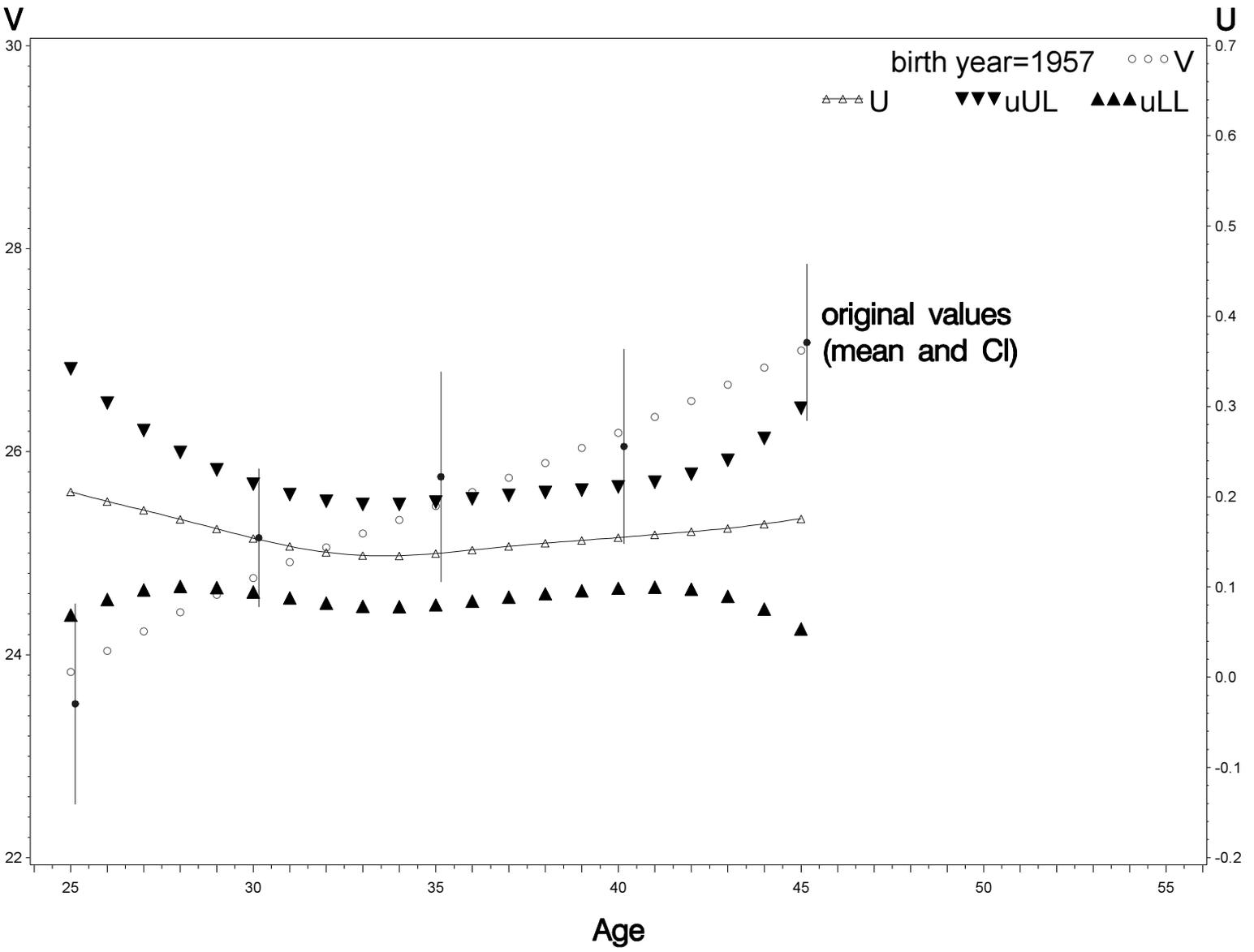}}
\end{center}
\caption{R(0.7,0.9) BMI, Men. C-trends (U) and estimates of means (V) by age for selected cohort.
\label{uv_co}}
\end{figure}

Figure~\ref{uv_co} displays original data ($mean \pm CI$), estimates of $v$ and estimates of C-trend with 95\% Confidence Interval along cohort line  for a selected cohort.
 
These figures illustrate the principle "one figure is better than
one hundred tables", though all the underlying data are available
and could be presented in a set of tables.
 
\begin{figure}[h]
\subfloat[Figure \ref{u}, R(0.7,0.7), $\mathnormal{R^2}=0.96$
]{\includegraphics[angle=90,width = 2.8in]{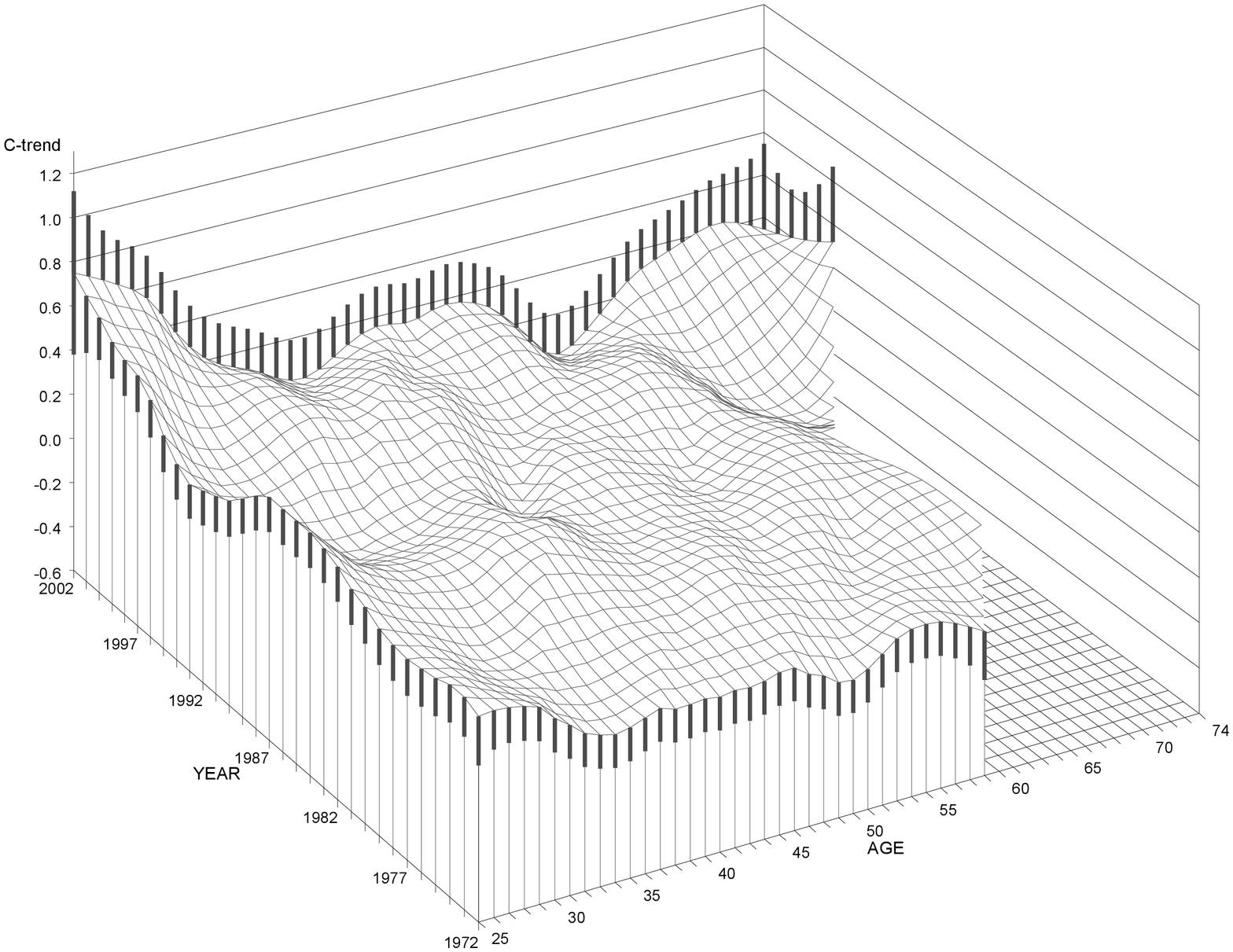}}
\subfloat[Figure \ref{u_CI}, R(0.7,0.7)]{\includegraphics[angle=90,width = 2.8in]{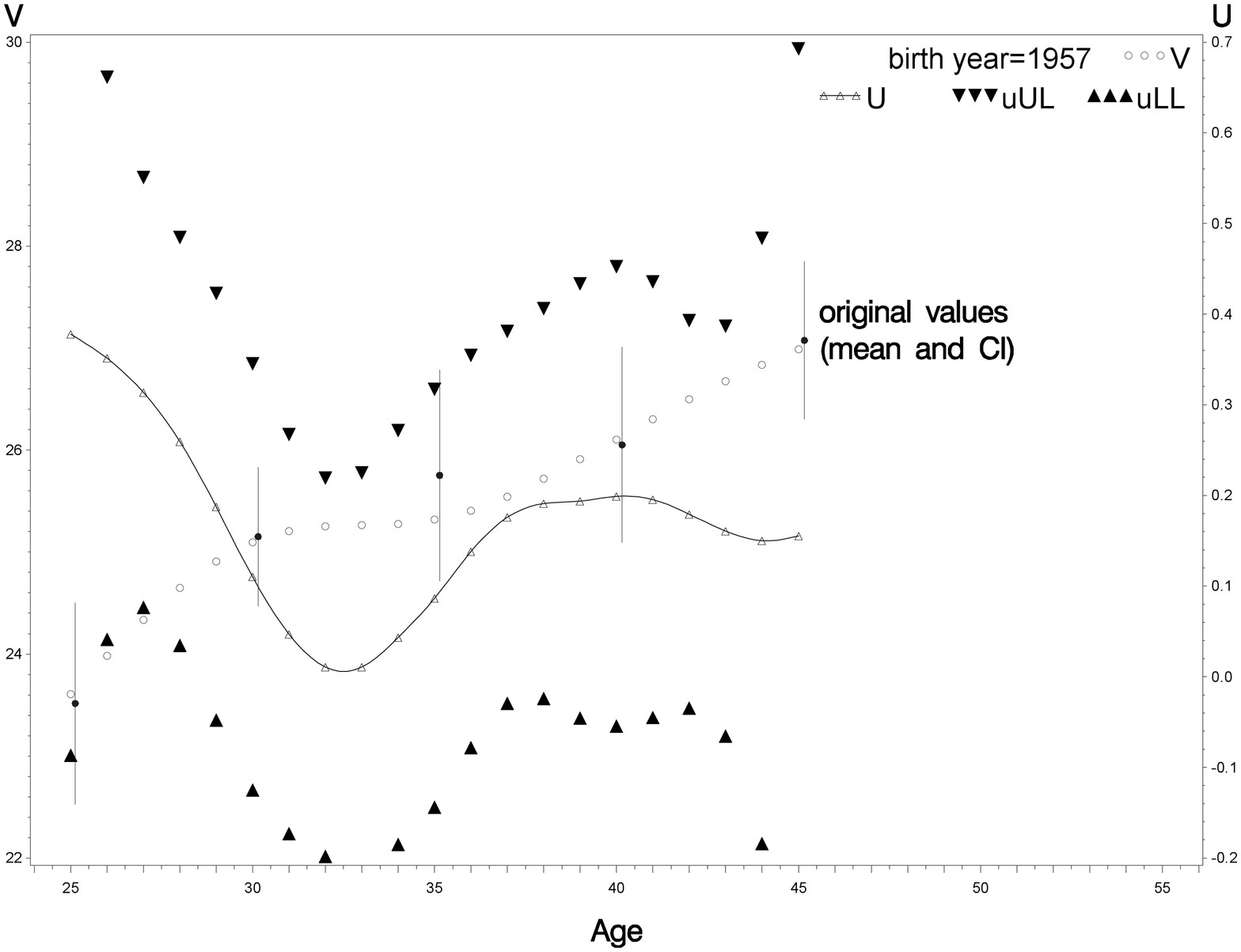}} \\
\subfloat[Figure \ref{u}, R(0.7,0.8), $\mathnormal{R^2}=0.94$
]{\includegraphics[angle=90,width = 2.8in]{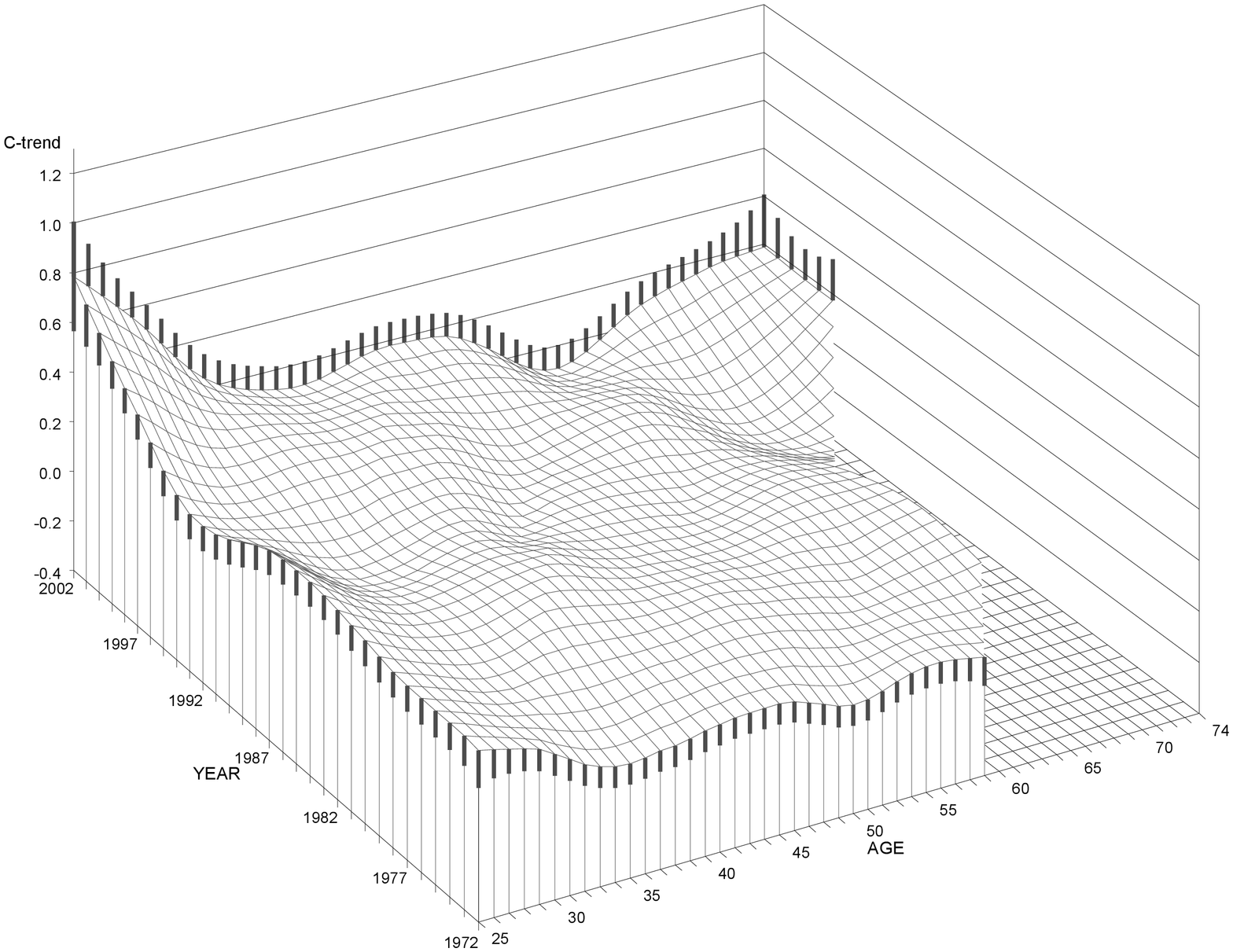}}
\subfloat[Figure \ref{u_CI}, R(0.7,0.8)]{\includegraphics[angle=90,width = 2.8in]{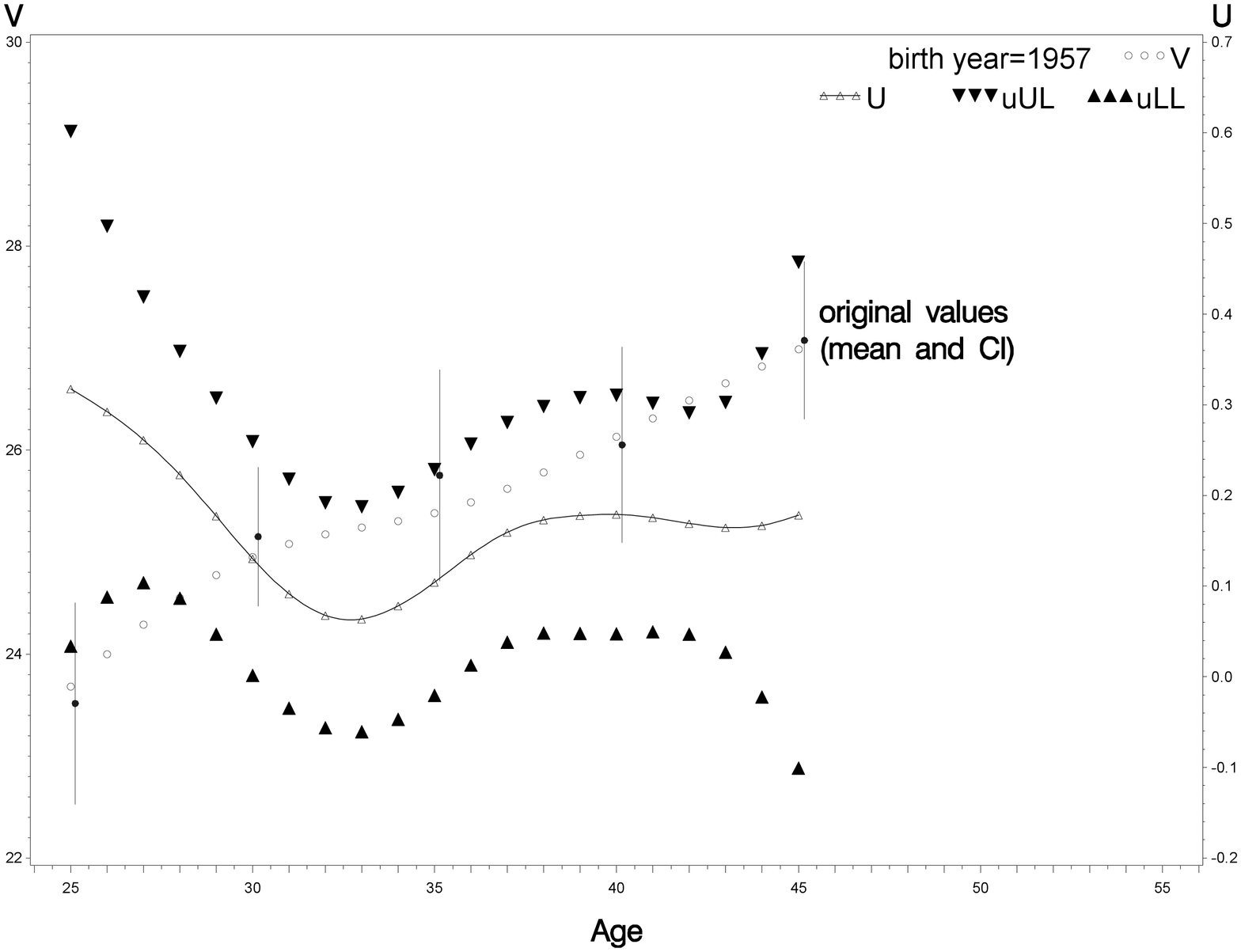}} \\
\subfloat[Figure \ref{u}, R(0.7,0.95), $\mathnormal{R^2}=0.91$
]{\includegraphics[angle=90,width = 2.8in]{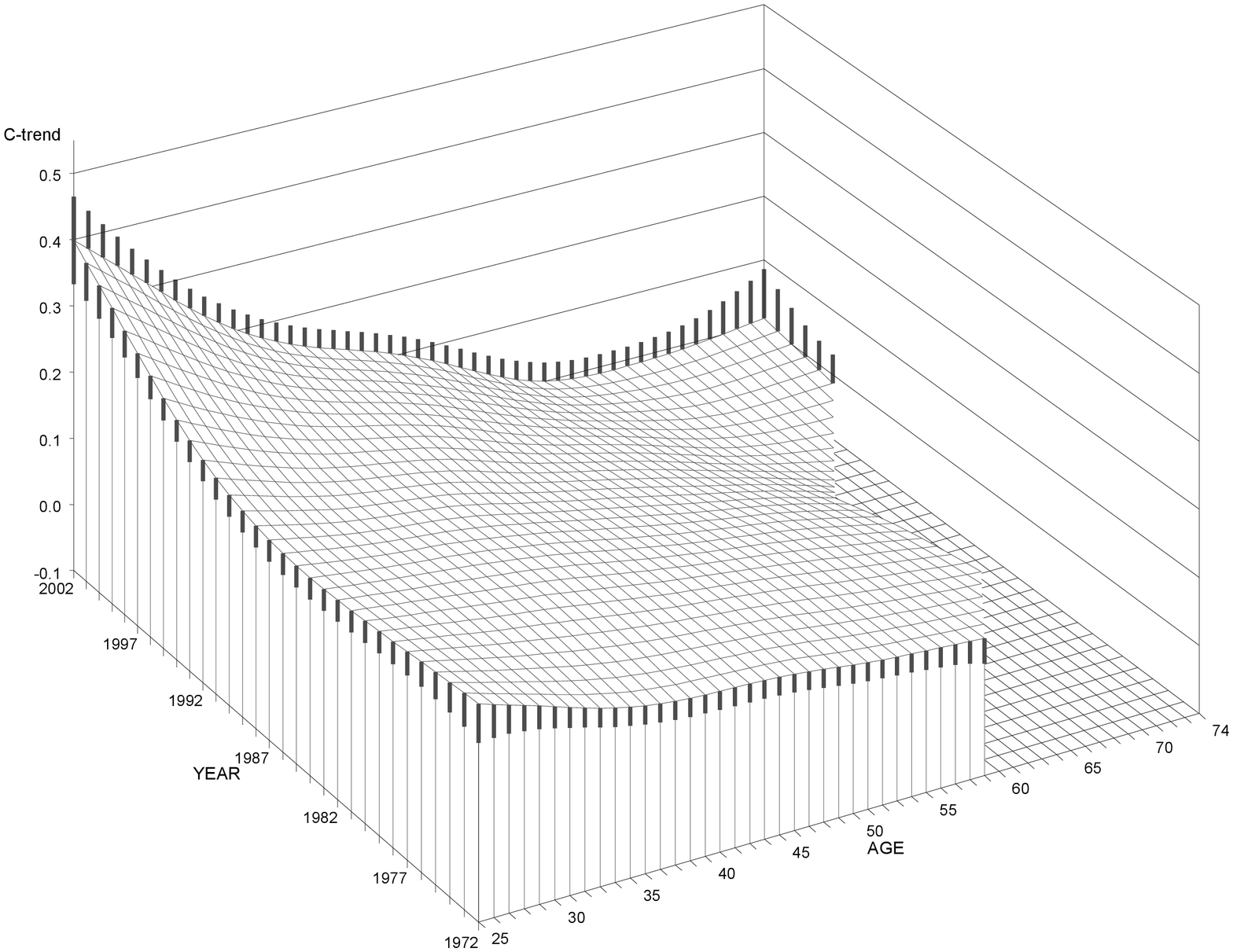}}
\subfloat[Figure \ref{u_CI}, R(0.7,0.95)]{\includegraphics[angle=90,width = 2.8in]{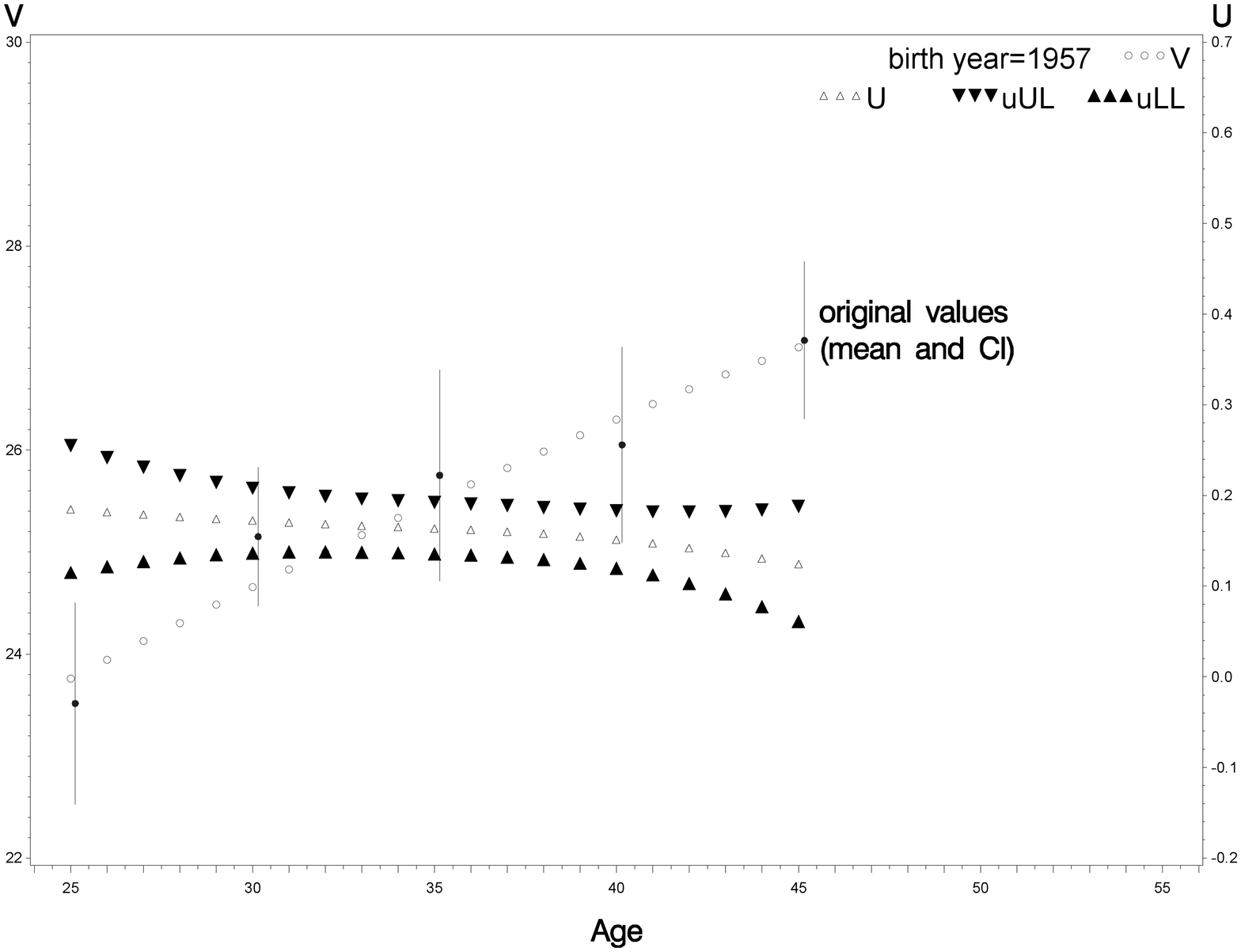}} \\
\caption{ Comparison of Figures \ref{u} and \ref{u_CI} by models R(0.7,0.7), R(0.7,0.8) and R(0.7,0.9) with parameter $\mathnormal{R^2}$ for Goodness of Fit }
\label{multifig}
\end{figure}

Figure~\ref{multifig}, along with  Figures ~\ref{Chart} and ~\ref{u_CI}, illustrates the choice of feasible solution.
 
Observe that with increasing $r_u$, the outcomes $R(0.7,0.7)$, $R(0.7,0.8)$,  $R(0.7,0.9)$ , $R(0.7,0.95)$ exhibit the following properties.
 
1. Confidence limits for C-trends estimates are decreasing  ( Figure~\ref{u_CI}), as well as maximal curvature of two-dimensional plot of C-trends estimates
over age and year.
 
2. The curvature of C-trends plot along cohort is decreasing ( Figure~\ref{u_CI} ). Relatively high  curvature may  occur due to a gap in the data (5 year gap between surveys) or due to odd values of some original  measurements. In that case  this plot exhibits non-monotonic function of age.
 
Thus, locally increased  curvature is a side effect of low $r_u$; a feasible solution with reasonable high   $(r_u=.9)$ is free of such  side affects nad less sensitive to possible biased original measurements. At the same time, the odd original values might be not a bias in measurements, rather it could result from  hight rates of migration-in or -out of population.
 
3. Goodness of fit $\mathnormal{R^2}=0.92$ is decreased from $0.96$ for $R(0.7,0.7)$    to $0.91$ for $R(0.7,0.9.5)$, remaining still significantly high.
 
Summing up, we may conclude that, as a feasible solution for the data available, the outcome   $R(0.7,0.9)$ could well be chosen. This is referred to as results in our further discussion.
 
\subsection{BMI dynamics: the main findings}
 
Figure~\ref{v} shows that mean BMI levels increase along cohort lines
throughout the study period, although they are different for
different birth cohort. Specific peaks and troughs   follow cohort
lines.
 
Figures ~\ref{u} ~\ref{u_CI} show that C-trends are decreased with age; they differ significantly with periods, especially for the age groups 25-29 and 30-34.
In particular, the alarmingly high C-trends are observed for the period 1997-2002, compared with period 1992-1997 for age groups 25-29,30-34 (significant difference is indicated in Figure~\ref{Chart}). In addition, the high rate of increasing C-trends  over year is observed for period 1992-2002  for age range 25-30.
 
The young generation coming to the observation frame exhibits rate of increasing BMI as high as 0.4 - 0.5 units per year. Which is about two times higher than in years 1972-1982. This is clear challenge to health management.
 
Recall that C-trends are proportional to  the external Driving Force (Modifier) which, in case of BMI, is the average hight-adjusted balance of calories.
Therefore the above findings induce the task for expert in social and economical areas to find out, why difference in calories consumed with food and burned  through physical activity is so alarmingly increased over time.

\section{Conclusion and Discussion \label{conclu}}
In this paper we have presented  a novel formulation of  the key principles of dynamic modeling  in application to health research, which justify the structure and interpretation of the core models dealing with C-trends.
 
In particular,
according to these principles, traditional  risk  factors' indicators  fall into
two categories, State Variables and Modifiers (see section \ref{ModPopLev} ),  having  different dynamical nature and, hence, playing different roles in the model and analysis.
 
As corollary of this, circular trends for State Variables have no sense at all. At the same time, only State Variables may determine instantaneous hazard rate of failure.
In dynamic models, causality is postulated: changes are due to Driving Forces (Modifiers), existing in the real world.
In case of consecutive survey data, C-trends are believed to be proxies for Driving Forces,  providing the  tool
for three main practical tasks: analysis, prediction and control of  health on population level ( see section \ref{ModPopLev})

We have used these principles as a framework for developing  the dynamic model  of simulating the temporal changes in characteristics of a real-world
object - population. In the course of this process, first, we
have identified two interacting objects, population and its
environment, on the top aggregation level. Further system analysis
has led us to breaking down the study population  into a set of
potentially infinitesimally narrow birth cohorts, carrying over time
health state profiles expressed in terms of health related
indicators (State Variables).
 
The model employs the \emph{health field} concept, suggesting
existence of  an influencing factors (Modifiers), generated by environment and acting
on the population, specific for each calendar year and age, and
causing within-cohort changes of the  health indicator with
rate of change corresponding to the strength of this factors.
 
For illustrative purposes we have selected one-parameter case with
continuous, normally distributed parameter and with strength
numerically equal to rate of change.  While keeping model reasonably
realistic, these simplifications help to highlight  the key
properties of the dynamic model of population health and method of
its identification - the Dynamic Regression Method.
 
In the illustrative example,  we have shown that the Dynamic
Regression Method applied to aggregated data, DRM3(A),   provides a sensible view on the BMI dynamics. It
reveals clear difference between dynamics the levels of the parameter and its
C-trends. From practical prospectives, it is C-trends, not levels,
which primarily seem to be modifiable by preventive activities or
involuntary changes affecting the population. It is worth noting
that outcomes from the DRM3(A)  analysis serve as data for the next-level
analysis, involving additional information and aiming at finding
reasonable explanation of the observed dynamics (diagnostic property
of DRM). One of the important complementary component for such an
analysis is dynamics of the population size (we have developed a
modification of the DRM for that type of data, this is a subject for
one of the next publication). If there is significant migration "in"
or "out" of the study population, the observed effects could be
entirely or partially due to the population instability (health
selective effect). The outcomes from the DRM analysis could be used
straightforwardly for prediction of the age-specific profile of the
State Variable, say, for 5 year  period, by applying the C-trends at
the last year of the study period to the estimates of the
parameter's levels at that year. Such a projection will not cover
the cohorts, not included in the study age range at the last study
year.
 
Recall that this method has been developed  as an alternative to the
secular trends and APC approach used so far. In this respect, it is worth noting that
the model presented here is characterized by local cohort trends
(C-trends), which have clear interpretation: changes in the State Variable of the same physical entity per time unit. If we will
formally calculate a characteristics resembling age-specific secular
trend, we will obtain  a difference between two different physical
entities (birth cohorts), caught occasionally at the moments of
measurement. Hence, it may behave quite arbitrarily. In other words,
in the view of the dynamic modeling approach,  secular trends do not
exist in nature. In one  special case only, when all the age
profiles of a State Variable are the same over calendar years
(stationary case), formally calculated secular trends will be equal
to zero at each age within the study age range. Only in that trivial
case, secular trends possess both, predictive and diagnostic power.
However, even in this case, secular trends are kind of statistical fallacy, due to  missing causality.
As to APC approach, the main methodological drawback of it is treating Age, Period and Cohort as linked algebraically,
while Cohort brings a differential component to the problem: namely,
derivative of the target variable along cohort line is a function of Age and Period.
This immediately removes the notorious conundrum of the APC method - linear dependency of the participating factors.
 
There are certain restrictions in using the current version of DRM3
methods,  imposed by
the size of the problem, due to using matrix operations.
Switching to Bayesian framework and employing Markov Chain Monte Carlo
methods \cite{Gel1} may solve these problems.
 
The  simplified dynamic equation used in the current model could  be
modified, accounting for the fact  that rate of change may  depend also on
the current level of the State Variable.
 
The more comprehensive model needs to be developed,
comprising multiple State Variables, and corresponding C-trends as a
linear functions of current State Variables.
 
Also, the model is to be developed processing measurements which are Modifiers, not Sate Variables, such as smoking and physical activity.
In some sense, this is an inverse problem to one presented here.
 
Collection of such models could be a
powerful practical tool for prediction of population health for about 5 year span.

\section*{Acknowledgements}
We thank The National FINRISK Study steering group for providing the data for illustrative
analysis.

\bibliography{v20_bib}

\begin{thebibliography}{}

\bibitem[\protect\astroncite{{Brown University}}{1995}]{BrownUni}
{Brown University} (1995).
\newblock {\em Learning Dynamical Systems: A Tutorial}
\newblock Available as
  {http://www.cs.brown.edu/research/ai/dynamics/tutorial/home.html}

\bibitem[\protect\astroncite{Chen et~al.}{2003}]{Chen2003}
Chen, X., Li, G., Unger, J.~B., Liu, X., and Johnson, C.~A. (2003).  Secular
  trends in adolescent never smoking from 1990 to 1999 in {C}alifornia: and
  age-period-cohort analysis.
\newblock {\em Am J Public Health}, 93:2099--104.

\bibitem[\protect\astroncite{DiLiberti and Lorenz}{2001}]{DiLiberti2001}
DiLiberti, J.~H. and Lorenz, R.~A. (2001).  Long-term trends in childhood
  diabetes mortality: 1968–1998.
\newblock {\em Diabetes Care}, 24:1348 -- 1352.

\bibitem[\protect\astroncite{Dobbin and Gatowski}{1999}]{JudgeDeskbook}
Dobbin, S.~A. and Gatowski, S.~I. (1999).
\newblock {\em Deskbook on the Basic Philosopies and Methods of Science,
  {C}hapter 4: Quantitative and Qualitative Research.}
\newblock Available as {http://www.unr.edu/bench/chap04.htm}

\bibitem[\protect\astroncite{Dobson et~al.}{1998a}]{Dobson1988a}
Dobson, A.~J., Evans, A., Ferrario, M., Kuulasmaa, M., Moltchanov, V., Sans,
  S., Tunstall-Pedoe, H., Tuomilehto, J., Wedel, H., and {Yarnell, J., for the
  WHO MONICA Project} (1998a).  Changes in estimated coronary risk in the
  1980s: data from 38 populations in the {WHO MONICA Project}.
\newblock {\em Ann Med}, 30:199 -- 205.

\bibitem[\protect\astroncite{Dobson et~al.}{1998b}]{Dobson1988b}
Dobson, A.~J., Kuulasmaa, K., Moltchanov, V., Evans, A., Fortmann, S.~P.,
  Jamrozik, K., Sans, S., and {Tuomilehto, J. for the WHO MONICA Project}
  (1998b).  Changes in cigarette smoking among adults in 35 populations in the
  mid-1980s.
\newblock {\em Tobacco Control}, 7:14--21.

\bibitem[\protect\astroncite{Gelman et~al.}{1995}]{Gel1}
Gelman, A., Carlin, J.~B., Stern, H.~S., and Rubin, D.~B. (1995).
\newblock {\em Bayesian Data Analysis.}
\newblock Chapman \& Hall, London

\bibitem[\protect\astroncite{Gregg et~al.}{2005}]{Edward2005}
Gregg, E.~W., Cheng, Y.~J., Cadwell, B.~L., Flegal, K.~M., Narayan, K. M.~V.,
  and Williamson, D.~F. (2005).  Secular trends in cardiovascular disease risk
  factors according to body mass index in {US} adults.
\newblock {\em JAMA}, 293:1868 -- 74.

\bibitem[\protect\astroncite{Hargrove}{1998}]{Harg}
Hargrove, J.~L. (1998).
\newblock {\em Dynamic Modeling in the Health Sciences.}
\newblock Modeling Dynamic Systems. Springer-Verlag, New York

\bibitem[\protect\astroncite{Havulinna}{2014}]{Havulinna}
Havulinna, A. (2014).  Bayesian age-period-cohort models with versatile
  interactions and long-term predictions: mortality and population in finland
  1878-2050.
\newblock {\em Statistics in Medicine}, 33:845--856. DOI:10.1002/sim.5954.

\bibitem[\protect\astroncite{Holford}{1991}]{Holford1991}
Holford, T.~R. (1991).  Understanding the effects of age, period, and cohort on
  incidence and mortality rates.
\newblock {\em Annu Rev Public Health}, 12:425 -- 457.

\bibitem[\protect\astroncite{{Jiang} et~al.}{2013}]{Jiang}
{Jiang}, T., {Gilthorpe MS}, {Shiely F}, {Harrington JM}, {Perry IJ}, {Kelleher
  CC}, and {Tu Y-K} (2013).  Age-period-cohort analysis for trends in body mass
  index in ireland.
\newblock {\em BMC Public Health}, 13:889.

\bibitem[\protect\astroncite{Kautiainen et~al.}{2002}]{Kautiainen2002}
Kautiainen, S., Rimpela, A.~H., Vikat, A., and Virtanen, S.~M. (2002).  Secular
  trends in overweight and obesity among finnish adolescents in 1977-1999.
\newblock {\em Int J Obes Relat Metab Disord}, 26:544 -- 52.

\bibitem[\protect\astroncite{Kuulasmaa et~al.}{2000}]{Kuulasmaa2000}
Kuulasmaa, K., Tunstall-Pedoe, H., Dobson, A., Fortmann, S., Sans, S., Tolonen,
  H., Evans, A., Ferrario, M., and {Tuomilehto, J. for the WHO MONICA Project}
  (2000).  Estimation of contribution of changes in classic risk factors to
  trends in coronary-event rates across the {WHO MONICA Project} populations.
\newblock {\em Lancet}, 355:675--87.

\bibitem[\protect\astroncite{Luenberger}{1979}]{Luenberger}
Luenberger, D.~G. (1979).
\newblock {\em Introduction to Dynamic Systems: Theory, Models, and
  Applications.}
\newblock John Wiley \& Sons, New York

\bibitem[\protect\astroncite{Moltchanov}{1993}]{Moltchanov1993}
Moltchanov, V. (1993).  The projection of the theory and methodology of the
  dynamic systems into epidemiological research.
\newblock {\em Can J Cardiol}, 9:88--89.

\bibitem[\protect\astroncite{Moltchanov}{2012}]{Moltchanov2012}
Moltchanov, V. (2012).  Dynamic modeling in health research as a framework for
  developing statistical applications free of misuse of statistics.
\newblock {\em arXiv:1211.1310v3 [stat.ME]}

\bibitem[\protect\astroncite{Moltchanov et~al.}{1999}]{Moltchanov1999}
Moltchanov, V., Kuulasmaa, K., and {Torppa, J. for the WHO MONICA Project}
  (1999).
\newblock {\em Quality assessment of demographic data in the {WHO MONICA
  Project}.}
\newblock Available as
  {http://www.ktl.fi/publications/monica/demoqa/demoqa.html}

\bibitem[\protect\astroncite{Moltchanov and Mik'halskii}{2008}]{MolMic}
Moltchanov, V.~A. and Mik'halskii, A.~I. (2008).  Estimation of dynamics of
  risk factors by the dynamic regression method.
\newblock {\em Automation and Remote Control}, 69:125--140.

\bibitem[\protect\astroncite{Neumaier}{1999}]{Neumaier}
Neumaier, A. (1999).  Solving ill-conditioned and singular linear systems: A
  tutorial on regularization.
\newblock {\em SIAM Review}, 40:636--666.

\bibitem[\protect\astroncite{Rao}{1973}]{Rao}
Rao, R.~C. (1973).
\newblock {\em Linear Statistical Inference and its Applications. Second
  edition.}
\newblock John Wiley \& Sons, New York

\bibitem[\protect\astroncite{{SAS Institute Inc.}}{2011a}]{SAS9.3PROC}
{SAS Institute Inc.} (2011a).
\newblock {\em Base SAS\textsuperscript{\textregistered} 9.3 Procedures Guide}
\newblock Cary, NC: SAS Institute Inc.

\bibitem[\protect\astroncite{{SAS Institute Inc.}}{2011b}]{SAS9.3IML}
{SAS Institute Inc.} (2011b).
\newblock {\em SAS/IML\textsuperscript{\textregistered} 9.3 User's Guide}
\newblock Cary, NC: SAS Institute Inc.

\bibitem[\protect\astroncite{{SAS Institute Inc.}}{2011c}]{SAS9.3STAT}
{SAS Institute Inc.} (2011c).
\newblock {\em SAS/STAT\textsuperscript{\textregistered} 9.3 User's Guide}
\newblock Cary, NC: SAS Institute Inc.

\bibitem[\protect\astroncite{{SAS Institute Inc.}}{2012a}]{SAS9.3GRAPH}
{SAS Institute Inc.} (2012a).
\newblock {\em SAS/GRAPH\textsuperscript{\textregistered} 9.3 User's Guide}
\newblock Cary, NC: SAS Institute Inc.

\bibitem[\protect\astroncite{{SAS Institute Inc.}}{2012b}]{SAS9.3L}
{SAS Institute Inc.} (2012b).
\newblock {\em SAS\textsuperscript{\textregistered} 9.3 Language Reference:
  Concepts}
\newblock Cary, NC: SAS Institute Inc.

\bibitem[\protect\astroncite{Selvin}{1996}]{Selvin1996}
Selvin, S. (1996).
\newblock {\em Cohort data: description and illustration.}
\newblock Oxford, England: Oxford University Press

\bibitem[\protect\astroncite{Tolonen et~al.}{2000}]{surveydb}
Tolonen, H., Kuulasmaa, K., and {Ruokokoski, E. for the WHO MONICA Project}
  (2000).
\newblock {\em MONICA population survey data book.}
\newblock Available as
  {http://www.ktl.fi/publications/monica/surveydb/title.htm}

\bibitem[\protect\astroncite{Tunstall-Pedoe et~al.}{2003}]{MoniMono}
Tunstall-Pedoe, H., editor. Prepared~by H~Tunstall-Pedoe, Kuulasmaa, K.,
  Tolonen, H., Davidson, M., and {Mendis, S. with 64 other contributors for The
  WHO MONICA Project} (2003).
\newblock {\em MONICA Monograph and Multimedia Sourcebook.}
\newblock Geneva: World Health Organization

\bibitem[\protect\astroncite{Yang and Land}{2013}]{Yang}
Yang, Y. and Land, K.~C. (2013).
\newblock {\em Age-Period-Cohort Analysis: New Models, Methods, and Empirical
  Applications}
\newblock Chapman \& Hall/CRC Interdisciplinary Statistics Chapman and Hall/CRC

\end{thebibliography}
\bibliographystyle{astronm}
 
\newpage
 
\newpage
 
 
\newpage
 
\begin{table}[!hbp]
\begin{center}
\caption[smallcaption]{The BMI analysis data for North Karelia, Finland, Men\\
Number of observations, sampling frame,  number of missing values  and survey periods \\
by survey year.}
\begin{tabular}{ccrc cccr rr}
& sampling &      &  \multicolumn{2}{c}{Age} & Part.rate &\multicolumn{2}{c}{BMI}  &\multicolumn{2}{c}{Period (month)}\\
year & frame    & Nobs & Min & Max                &(\%) & miss. (\%)  & N.used   & Start & Finish  \\\toprule
\hline
1972 & 1 & 2657 & 25 & 59 & 94 & 6.2 & 2492 & 2 & 9 \\
1977 & 1 & 2980 & 25 & 64 & 87 & 0.9 & 2953 & 1 & 4 \\
1982 & 2 & 1538 & 25 & 64 & 76 & 0.1 & 1537 & 1 & 4 \\
1987 & 2 & 1561 & 25 & 64 & 79 & 5.1 & 1481 & 1 & 4 \\
1992 & 2 & 673 & 25 & 64 & 68 & 0.0 & 673 & 1 & 3 \\
1997 & 2 & 1171 & 25 & 74 & 72 & 6.1 & 1100 & 1 & 6 \\
2002 & 2 & 863 & 25 & 74 & 66 & 6.3 & 809 & 1 & 4 \\
 
\hline
All & 1,2 & 11443 & 25 & 74 & 66-94  & 3.5 & 11045 & 1 & 9 \\
\hline
\end{tabular}
\small
\item Sampling frame: 1=simple random, 2= stratified by 10 years age groups.
\label{tbl:data}

\end{center}
 
\end{table}

\end{document}